\documentclass[10pt]{article}
\usepackage[T1]{fontenc}
\usepackage{lmodern}
\usepackage[a4paper,margin=1in]{geometry}
\usepackage{amsmath,amssymb,amsthm,mathtools}
\usepackage{microtype}
\usepackage[skip=0.45\baselineskip plus 2pt,indent=0pt]{parskip}
\usepackage{graphicx,booktabs,array,float,flafter}
\usepackage{titletoc}
\usepackage{etoolbox}
\usepackage{aliascnt}
\usepackage[colorlinks=true,linkcolor=blue,citecolor=blue,urlcolor=blue]{hyperref}
\usepackage[nameinlink,noabbrev,capitalize]{cleveref}
\newif\ifstandalonesupplement
\expandafter\ifstrequal\expandafter{\jobname}{supplement}{\standalonesupplementtrue}{}
\newcommand*{\ghzdatarevision}{1fa53f858121889bf3fef00c86e52ab27573a257}
\newcommand*{\ghzdatafile}[2]{\href{https://github.com/xliu2022/xliu2022.github.io/blob/\ghzdatarevision/GHZ_supplementary_data/#1}{#2}}
\newcommand*{\ghzdataarchive}{\href{https://raw.githubusercontent.com/xliu2022/xliu2022.github.io/\ghzdatarevision/GHZ_supplementary_data/supplementary_data.zip}{\nolinkurl{supplementary_data.zip}}}
\titlecontents{suppsection}[2.25em]{\addvspace{2pt}\bfseries}
  {\contentslabel{2.25em}}{}{\hfill\contentspage}
\apptocmd{\thebibliography}{\addcontentsline{toc}{section}{\refname}}{}{}
\numberwithin{equation}{section}
\newtheorem{theorem}{Theorem}[section]
\newaliascnt{lemma}{theorem}
\newtheorem{lemma}[lemma]{Lemma}
\aliascntresetthe{lemma}
\newaliascnt{proposition}{theorem}
\newtheorem{proposition}[proposition]{Proposition}
\aliascntresetthe{proposition}
\newaliascnt{corollary}{theorem}
\newtheorem{corollary}[corollary]{Corollary}
\aliascntresetthe{corollary}
\newaliascnt{claim}{theorem}
\newtheorem{claim}[claim]{Claim}
\aliascntresetthe{claim}
\theoremstyle{definition}

\crefname{claim}{Claim}{Claims}
\Crefname{claim}{Claim}{Claims}
\crefname{section}{Section}{Sections}
\Crefname{section}{Section}{Sections}
\crefname{equation}{}{}
\Crefname{equation}{Equation}{Equations}
\allowdisplaybreaks[1]

\hypersetup{
 pdftitle={Unbounded GHZ dimension from fixed-order multiphoton sources},
 pdfauthor={Heng Li and Xizhi Liu},
 pdfsubject={Fixed source order, unbounded GHZ dimension, rigid hypergraphs}
}
\makeatletter
\patchcmd{\@maketitle}{\null\vskip 2em}{}{}{}
\makeatother
\title{\Large\bfseries Unbounded GHZ dimension from fixed-order multiphoton sources}
\author{
Heng~Li\thanks{School of Mathematics, Shandong University, Jinan, China, and Extremal Combinatorics and Probability Group (ECOPRO), Institute for Basic Science (IBS), Daejeon, South Korea. \href{mailto:heng.li@sdu.edu.cn}{\nolinkurl{heng.li@sdu.edu.cn}}.}
\and
Xizhi~Liu\thanks{School of Mathematical Sciences, University of Science and Technology of China, Hefei, China. \href{mailto:liuxizhi@ustc.edu.cn}{\nolinkurl{liuxizhi@ustc.edu.cn}}.}
}
\date{\today}
\ifstandalonesupplement
\usepackage{xr}
\newcommand*{\usepaperlabel}[1]{%
  \csletcs{r@#1}{r@paper-#1}%
  \csletcs{r@#1@cref}{r@paper-#1@cref}}
\forcsvlist{\usepaperlabel}{sec:third-part,tab:finite-thirteen,tab:twelve-networks}
\hypersetup{pdftitle={Supplemental Material for Unbounded GHZ dimension from fixed-order multiphoton sources}}
\title{\large Supplemental Material for\\[2pt]\Large\bfseries Unbounded GHZ dimension from fixed-order multiphoton sources}
\date{}
\fi

\begin{document}
\maketitle
\ifstandalonesupplement\else
\begin{abstract}
We study the local dimension of Greenberger--Horne--Zeilinger (GHZ) states obtained by selecting events with one photon per output path. Each coherent source emits photons with a fixed internal label into a distinct set of paths, and the sets for each label partition the outputs. For two-photon sources, the maximum local dimension is 2 for every even number of paths greater than 4. However, we show that three-photon sources can produce GHZ states of arbitrarily large local dimension. To prove this, we view the three paths fed by each source as an edge of a $3$-uniform hypergraph, so that accepted emission patterns correspond exactly to perfect matchings. We construct such hypergraphs with arbitrarily many edge-disjoint perfect matchings, one per label, and no others. The construction extends to $r$-uniform hypergraphs for every fixed $r\ge3$, yielding a GHZ dimension that grows at least polynomially as the number of paths ranges over the positive multiples of $r$.
All proofs were generated by GPT.
\end{abstract}

\section{Introduction}\label{sec:introduction}

Krenn et al.~\cite{KrennEtAl2026NoGo} asked whether three-photon sources can produce Greenberger--Horne--Zeilinger (GHZ) states of arbitrarily large local dimension. A source emits one photon into each of $r$ distinct paths, and all $r$ photons carry its label. We retain events with one photon in each of the $n$ output paths. The $d$-level generalization~\cite{Lawrence2014} of the qubit GHZ state~\cite{GreenbergerHorneZeilinger1989,GreenbergerHorneShimonyZeilinger1990} is
\begin{equation}
 |\mathrm{GHZ}_{n,d}\rangle\coloneqq\frac{1}{\sqrt d}\sum_{i=1}^{d}|i\rangle^{\otimes n}.
 \label{eq:GHZ-definition}
\end{equation}
In the $i$th term all $n$ photons carry label $i$. Polarization GHZ states with up to 12 photons have been observed at $d=2$~\cite{BouwmeesterEtAl1999,ZhaoEtAl2004,WangEtAl2016,ZhongEtAl2018}. Erhard et al.~\cite{ErhardMalikKrennZeilinger2018} reached $d=3$ with 3 photons using a multiport suggested by the Melvin algorithm~\cite{KrennEtAl2016}.

Krenn, Gu, and Zeilinger~\cite{KrennGuZeilinger2017} represented output paths by vertices and two-photon sources by graph edges. In our model distinct sources feed distinct sets of paths. The sources of each label cover every path exactly once, so their edges form a perfect matching. Any additional perfect matching would give an accepted emission pattern in which at least two photons carry different labels. Bogdanov's graph theorem~\cite{Bogdanov2017} gives a maximum local dimension of 3 on 4 paths and 2 for every even $n\ge6$.

Gu, Chen, and Krenn~\cite{GuChenKrenn2020} extended the model to hypergraphs, with each source represented by the set of paths it feeds. Their construction on 6 paths uses all 20 triples. Each triple has a unique partner covering the remaining 3 paths. Giving each of these 10 pairs its own label yields local dimension 10. On 9 paths a given triple has 10 possible completions to a perfect matching. We must select a family of triples in which each triple has exactly one completion. For each fixed $r\ge3$, we construct networks with arbitrarily many labels in which every perfect matching uses a single label.

\begin{theorem}\label{thm:ghz-growth}
For each integer $r\ge3$, there is a constant $c_r>0$ with the following property. For every positive multiple $n$ of $r$, there is a network of coherent $r$-photon sources whose normalized state after postselection on one photon per output path is $|\mathrm{GHZ}_{n,d}\rangle$ with $d\ge c_r n^{2(r-2)/7}$. The $n$ paths are partitioned into $r$ groups of equal size, and each emitting source sends one photon to each group. All sources can have the same emission probability, with the vacuum and emission amplitudes of each source in phase.
\end{theorem}

For fixed $r$, a deterministic algorithm lists the output paths and internal label of each source in time polynomial in $n$. In the lattice constructions we scan boxes of dimension at most 3 and side lengths $O_r(n)$ in lexicographic order and form all tuples from $r-2$ lists of at most $n^2$ matchings each.

In the triplet construction, sources with different labels can together cover every path in the first two groups exactly once. We assign each source a third path so that every such mixed emission pattern misses a path in the third group (see \cref{fig:comb}). For $r>3$, we combine $r-2$ triplet networks that share the first two groups. The perfect matchings of the resulting network correspond exactly to choices of one matching from each triplet network.

For fixed $r$, every network of $r$-photon sources that realizes a GHZ state in this model satisfies $d=O_r(n^{r-1}/\log n)$. Gu, Chen, and Krenn~\cite[Sec.~III\,B]{GuChenKrenn2020} reported that dimension 13 is attainable on 9 paths. We prove by elementary counting that 13 is optimal for unrestricted triples. If the paths are divided into three groups of 3 and every triple meets all three groups, the maximum is 6. The underlying hypergraph at dimension 6 is then unique up to relabeling within each group.

The proofs of the growth bound and the unrestricted nine-path upper bound are in Appendices~\ref{app:construction} and~\ref{app:finite}, respectively.

Both Palomar entries use the title and abstract of the \href{https://github.com/xliu2022/GHZ-unbounded-lean}{Lean~4 project}. Of the 20 theorems summarized there, \href{https://palomar-registry.org/entry.html?id=PALOMAR-2026-11-24-000007&version=1}{entry~000007} verifies nine on perfect matchings. These give the polynomial lower bound and the construction for a prescribed dimension (\cref{thm:growth,cor:prescribed}), as well as the harmonic and logarithmic upper bounds in \cref{sec:harmonic}. \href{https://palomar-registry.org/entry.html?id=PALOMAR-2026-11-24-000010&version=1}{Entry~000010} verifies seven results for coherent sources, including the resulting GHZ dimension bound (\cref{thm:ghz-growth}), source normalization, the postselection criterion and local phase correction (\cref{prop:source-rigidity}), and the optimal success probability for equal branch weights (\cref{prop:balanced-pump}). A \href{https://github.com/xliu2022/GHZ-unbounded-lean/blob/92a4ec9870602cc466a346e97edef26736ce21c4/comparator-finite.json}{separate list of four theorems} covers the unrestricted six-path maximum and the nine-path results in \cref{thm:finite-exact}. Lean proves the unrestricted nine-path upper bound by checking certificates that rule out dimension~14. The \href{https://github.com/xliu2022/GHZ-unbounded-lean/blob/92a4ec9870602cc466a346e97edef26736ce21c4/COVERAGE_AUDIT.md}{proof index} also covers grafting, cyclic lifts, and thresholds.

\section{The source model and rigid hypergraphs}\label{sec:model}

A source that emits into $r$ distinct paths specifies an $r$-element set of vertices. The emitting sources place one photon in every path exactly when their supports partition the vertex set. We use this correspondence to compute the amplitude of each accepted string of internal labels.

\subsection{The combinatorial problem}

For an integer $r\ge2$, an \emph{$r$-uniform hypergraph}, or \emph{$r$-graph}, is a collection of $r$-subsets of a finite vertex set. All hypergraphs here are simple. We identify $H$ with its edge set and denote its vertex set by $V(H)$. For positive integers $q$, write $[q]\coloneqq\{1,\ldots,q\}$. A \emph{perfect matching} is a family of pairwise vertex-disjoint edges that covers $V(H)$. It has $m$ edges when $|V(H)|=rm$.

An $r$-graph is \emph{$d$-rigid} if its edge set is the disjoint union of $d$ perfect matchings $M_1,\ldots,M_d$ and these are its only perfect matchings. For $r\mid n$, let $f_r(n)$ be the largest $d$ for which there is a $d$-rigid $r$-graph on $n$ vertices. In a \href{https://mathoverflow.net/questions/347793/hypergraphs-with-only-disjoint-perfect-matchings}{2019 MathOverflow question}, Krenn asked for $f_r(rm)$ for all $r,m\ge2$. An $r$-graph is \emph{balanced $r$-partite} if its vertices can be partitioned into $r$ parts of equal size with each edge meeting every part once. Let $g_r(m)$ be the largest $d$ for a $d$-rigid $r$-graph of this kind with $m$ vertices in each part. Equivalently, in either class one may maximize the number of perfect matchings in a hypergraph whose perfect matchings are pairwise edge-disjoint. Deleting every edge that belongs to no perfect matching leaves precisely the union of those matchings.

In particular, $g_r(m)\le f_r(rm)$. Every vertex of a $d$-rigid hypergraph belongs to one edge of each $M_i$, so the hypergraph is $d$-regular. Consequently,
\[
 |H|=dm=\frac{dn}{r},
 \qquad
 f_r(n)\le\binom{n-1}{r-1},
 \qquad
 g_r(m)\le m^{r-1}.
\]
The union of any nonempty subfamily of the prescribed matchings has exactly those perfect matchings.

For $n=r$, a simple $r$-graph has at most one edge, so $f_r(r)=1$. For $n=2r$, every edge has a unique possible partner in a perfect matching, namely its complement. The complete $r$-graph therefore attains $f_r(2r)=\frac12\binom{2r}{r}$. For example, $f_3(6)=10$, as used in the construction of Gu, Chen, and Krenn~\cite{GuChenKrenn2020}. Now partition the $2r$ paths into $r$ groups of size 2. The $2^r$ edges meeting each group once fall into $2^{r-1}$ complementary pairs. Thus $g_r(2)=2^{r-1}$, giving $g_3(2)=4$ on the same 6 paths. How large can $f_r(rm)$ and $g_r(m)$ be as $m$ grows with $r$ fixed?

Bogdanov~\cite{Bogdanov2017} proved that $K_4$ is the only simple graph with at least 3 perfect matchings in which all perfect matchings are pairwise edge-disjoint. An even cycle has exactly 2 perfect matchings, so $f_2(4)=3$ and $f_2(n)=2$ for every even $n\ge6$. Chandran and Gajjala~\cite{ChandranGajjala2026} studied edge colorings in which every perfect matching is monochromatic. They characterized the graphs for which the maximum number of colors represented by perfect matchings is 1 or 2.

For fixed matching size $m\ge2$ and every $r\ge2$, Pohoata, Sauermann, and Zakharov~\cite[Theorem~4.3 of the preprint]{PSZ2025} construct a balanced $r$-partite $r$-graph on $rm$ vertices with at least $m^{r-D_m\sqrt r}$ prescribed perfect matchings, where $D_m$ depends only on $m$. Whenever $m$ disjoint edges are chosen from these matchings, all their matching indices must agree. This condition rules out shared edges and additional perfect matchings in their union.

Aharoni and Berger~\cite{AharoniBerger2009} asked how many $m$-edge matchings in an $r$-partite $r$-graph force a rainbow matching of size $m$, with one edge from each of $m$ distinct matchings. Alon~\cite{AlonRainbow2011} and Glebov, Sudakov, and Szab{\'o}~\cite{GlebovSudakovSzabo2014} proved upper bounds. Pohoata, Sauermann, and Zakharov~\cite[Theorem~1.2]{PSZ2025} proved that for fixed $r\ge3$ the threshold has order $m^r$ for both $r$-partite and general $r$-graphs. The hypergraphs in \cref{thm:growth} have no perfect matching with mixed indices, even when some indices repeat.

\subsection{A normalized source model}
\label{sec:optical-model}

The internal labels denote orthogonal states on each path, such as orbital angular momentum modes~\cite{MairEtAl2001,DadaEtAl2011}, time bins~\cite{BrendelEtAl1999,DeRiedmattenEtAl2002}, or frequency bins~\cite{OlislagerEtAl2010,KuesEtAl2017}.

Let $H$ be the union of edge-disjoint perfect matchings $M_1,\ldots,M_d$ on a set $V$ of $n=rm$ paths. Every photon emitted by a source in $M_i$ carries internal label $i\in[d]$. Give every edge of $M_i$ color $i$. The figures and exported source lists number the labels from $0$ to $d-1$. Since $M_i$ covers every path once, each mode $(v,i)$ belongs to exactly one source.

Cascaded down-conversion has produced photon triplets~\cite{HubelEtAl2010}, including polarization-entangled triplets~\cite{HamelEtAl2014}. In the model, the source for $e\in M_i$ prepares a coherent superposition of the vacuum $|0_e\rangle$ and the state $|1_e\rangle$ with one photon in each of the $r$ modes $(v,i)$, $v\in e$. Let $p_e\in(0,1)$ be the emission probability. With a common phase reference, take the vacuum amplitude of each source to be real and positive and write $\phi_e$ for the phase of its emission amplitude. The normalized joint state of the independent sources is
\begin{equation}
 |\Psi\rangle\coloneqq\bigotimes_{e\in H}\left(\sqrt{1-p_e}\,|0_e\rangle+e^{\mathrm{i}\phi_e}\sqrt{p_e}\,|1_e\rangle\right).
 \label{eq:normalized-source-state}
\end{equation}
Coherence between the accepted branches requires that neither the pump nor any unobserved mode retain information about which sources emitted. Aligning the paths of undetected photons can erase information about which source emitted~\cite{ZouWangMandel1991}. This principle underlies entanglement by path identity~\cite{KrennHochrainerLahiriZeilinger2017,Lahiri2018}, which Kysela et al.~\cite{KyselaEtAl2020} demonstrated for high-dimensional photon pairs.

Let $\Pi_1$ project onto the subspace with exactly one photon of arbitrary internal label in each spatial path. For a perfect matching $M$ of $H$, write $i_M(v)$ for the color of its edge through $v$. Summing the accepted terms of \cref{eq:normalized-source-state} over all perfect matchings $M$ of $H$ gives
\begin{align}
 \Pi_1|\Psi\rangle
 &=\sum_M B_M\bigotimes_{v\in V}|i_M(v)\rangle,\label{eq:matching-state}\\
 B_M&\coloneqq\prod_{e\in M}e^{\mathrm{i}\phi_e}\sqrt{p_e}\prod_{e\in H\setminus M}\sqrt{1-p_e}.\label{eq:matching-amplitude}
\end{align}
\begin{samepage}
\begin{proposition}\label{prop:source-rigidity}
Let $H$ be the union of $d$ edge-disjoint perfect matchings $M_1,\ldots,M_d$ on $n$ paths. Associate with each edge $e\in M_i$ a source as in \cref{eq:normalized-source-state}, with $0<p_e<1$ and label $i$ on every path of $e$. The set of label strings with nonzero amplitude in $\Pi_1|\Psi\rangle$ equals $\{(i,\ldots,i)\colon i\in[d]\}$ if and only if $H$ is $d$-rigid. In this case, if the branch amplitudes $B_{M_i}$ have equal magnitude, the normalized postselected state is $|\mathrm{GHZ}_{n,d}\rangle$ up to a diagonal unitary on one output path.
\end{proposition}
\end{samepage}
\begin{proof}
At each path $v$, the local label identifies the unique incident edge of that color. Hence the label string $(i_M(v))_{v\in V}$ determines $M$. Each prescribed matching gives the same label on every path. Any other perfect matching gives a distinct, nonconstant label string. All coefficients in \cref{eq:matching-amplitude} are nonzero. Thus the constant strings account for the entire projected state precisely when $H$ is $d$-rigid. The normalized coefficients are then $B_{M_i}/(\sum_j|B_{M_j}|^2)^{1/2}$. A diagonal unitary on one path removes their phases. If their moduli agree, each coefficient becomes $1/\sqrt d$.
\end{proof}

The single-photon Schmidt ranks used to classify multipartite entanglement~\cite{HuberDeVicente2013} can differ, as in the three-photon state of Malik et al.~\cite{MalikEtAl2016} with ranks $(3,3,2)$. In \cref{eq:GHZ-definition} the $d$ product states have pairwise orthogonal factors on both sides of every nontrivial bipartition, so the Schmidt rank across each cut is $d$.

In a more general source model, distinct perfect matchings can give the same label string~\cite{KrennGuSoltesz2019,GuErhardZeilingerKrenn2019}. Their amplitudes then add and can interfere, permitting other entangled states~\cite{GuChenZeilingerKrenn2019}. Optical networks for a prescribed state have been found by optimizing graph weights~\cite{KrennEtAl2021Theseus,RuizGonzalezEtAl2023} and by encoding the allowed matchings as Boolean constraints~\cite{CerveraLiertaEtAl2022}.

In the complex-weighted graph model the two ends of an edge may carry different labels. For simple connected graphs, Chandran and Gajjala bounded the GHZ dimension in terms of the number of vertices~\cite{ChandranGajjala2024} and the vertex connectivity~\cite{ChandranGajjalaConnectivity2026}. For $n>4$, Chandran, Gajjala, and Illickan~\cite{ChandranGajjalaIllickan2024} proved $d\le2$ for cubic graphs and for graphs of vertex connectivity at most 2. For even $n\ge4$, Krenn et al.~\cite{KrennEtAl2026NoGo} ruled out $d\ge n$ in this model. Rajasekaran~\cite{Rajasekaran2026Lean} gave a Lean proof excluding $d\ge3$ for even $n\ge6$.

Heralding with a Fourier interferometer~\cite{PaesaniEtAl2021} and sequential emission from an array of matter qubits~\cite{BellEtAl2022} give other protocols for GHZ states of arbitrary local dimension. With ideal destructive photon counting, we retain records with one count in each output path. In the absence of loss, repeated emission by a source puts at least two photons in each of its paths and is rejected. \Cref{fig:dimension-transition} compares the six-path examples and shows how each choice of a row and an exit specifies a projected perfect matching. \Cref{fig:comb} shows a choice of triples from two matchings that leaves a path in the third part uncovered.

\begin{figure}[htbp]
\centering
\includegraphics[width=\linewidth]{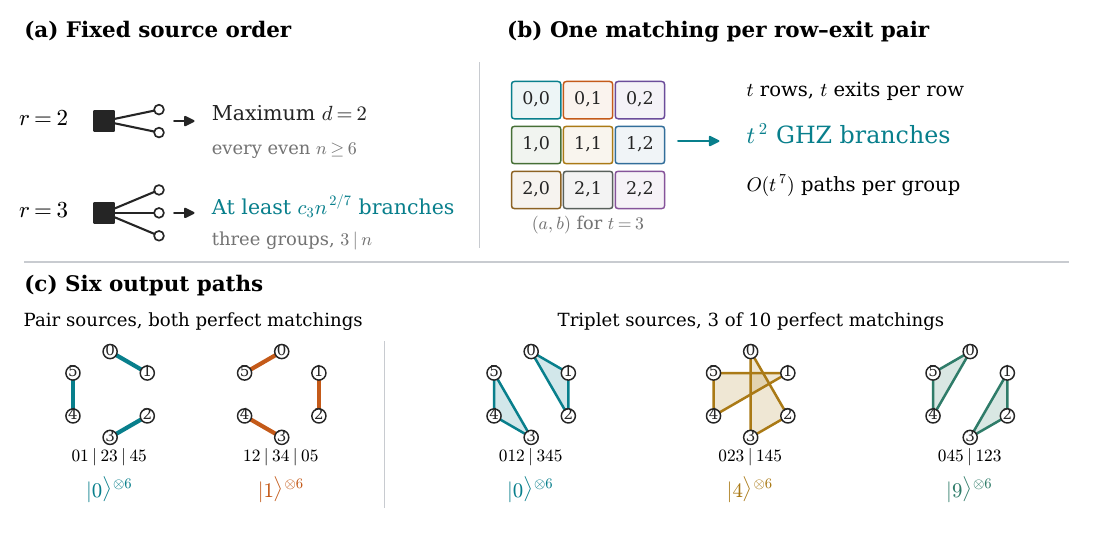}
\caption{Perfect matchings at fixed source order. (a) Maximum local dimension for pair sources and a lower bound for triplet sources in the model of \cref{sec:model}. (b) Each cell specifies a projected perfect matching by its row and exit in the double comb of \cref{sec:asymptotic}. The grid shows all nine choices for $t=3$. (c) Both perfect matchings of a 6-cycle and three of the ten complementary pairs of triples in the six-path construction of Gu, Chen, and Krenn~\cite{GuChenKrenn2020}. The numbers on the vertices denote output paths.}

\label{fig:dimension-transition}
\end{figure}

\section{A polynomial lower bound at fixed source order}\label{sec:asymptotic}

The following bound gives an explicit choice of $c_r$ in \cref{thm:ghz-growth}.

\begin{theorem}\label{thm:growth}
For all integers $r\ge3$ and $m\ge1$,
\[
 f_r(rm)\ge g_r(m)\ge\left(\frac{m}{128(r-2)^3}\right)^{2(r-2)/7}.
\]
\end{theorem}

Taking $m=n/r$ in \cref{thm:growth} and applying \cref{prop:source-rigidity} gives \cref{thm:ghz-growth} with $c_r=[128r(r-2)^3]^{-2(r-2)/7}$. In particular, the same construction gives $g_3(n/3)\ge(n/384)^{2/7}$ for every positive multiple $n$ of 3.

By retaining exactly $d$ of the matchings, we can prescribe the local dimension.

\begin{corollary}\label{cor:prescribed}
For integers $r\ge3$ and $d\ge2$, put $k=r-2$ and $m=\lceil128k^3d^{7/(2k)}\rceil$. There is a balanced $r$-partite $r$-graph on $n=rm$ vertices that is $d$-rigid and has $dm$ edges. In particular, $n=O_r(d^{7/(2(r-2))})$ and the number of source configurations is $O_r(d^{1+7/(2(r-2))})$.

\end{corollary}
\begin{proof}
By \cref{thm:growth}, this choice of $m$ gives at least $d$ prescribed perfect matchings. Retain any $d$ of them. Their union has exactly $d$ perfect matchings and $dm$ edges. For any $m'\ge m$, the same argument gives exactly $d$ perfect matchings and $dm'$ edges at part size $m'$.
\end{proof}

The balanced tripartite example in \cref{tab:twelve-networks} attains dimension 12 on 39 paths.

\subsection{How the third part rules out additional perfect matchings}\label{sec:third-part}

A bipartite perfect matching can be represented as a directed cycle cover. Let the two parts $A$ and $B$ be copies of the same set. The arc $u\to v$ represents the pair $(u_A,v_B)$. Let $\mathcal P_1,\ldots,\mathcal P_q$ be all the cycle covers of the chosen digraph. For each occurrence of an arc in a cover, we choose a vertex in the third part $C$.

For an arc $e$ of cover $\mathcal P_i$, the \emph{tag} $\mathbf z_e+\mathbf v_i$ specifies the third output path of the corresponding triple. The anchor $\mathbf z_e$ depends only on the arc, and the translation $\mathbf v_i$ depends only on the cover. Each of the 3 photons carries the internal label $i$. We choose the vectors so that each prescribed matching covers $C$ exactly once.

In the 9-vertex example of \cref{fig:comb}, the projections of $M_0$ and $M_1$ are the cycle covers $\mathcal P_0=\{uv,vu,ww\}$ and $\mathcal P_1=\{uv,vw,wu\}$, where $uv$ denotes $u\to v$. The vertices $c_0,c_1,c_2$ are the three spatial paths in $C$. Take scalar translations $v_0=0$, $v_1=1$ and anchors $z_{uv}=0$, $z_{vu}=z_{wu}=1$, $z_{ww}=2$, $z_{vw}=-1$. In $M_i$, the triple that projects to $e$ uses the third path $c_{z_e+v_i}$, as shown in panels (c) and (d).

The two cycle covers share only the arc $u\to v$. In $M_0$, replacing the triple that projects to this arc by its counterpart in $M_1$ leaves every path in $A$ and $B$ covered once. In $C$ the new choice covers $c_1$ twice and misses $c_0$. The reverse switch in $M_1$ covers $c_0$ twice and misses $c_1$. These are the only choices of triples from both matchings that cover $A$ and $B$ exactly once. Neither covers $C$ exactly once, so $M_0$ and $M_1$ are the only perfect matchings.

\begin{figure}[htbp]
\centering
\includegraphics[width=\linewidth]{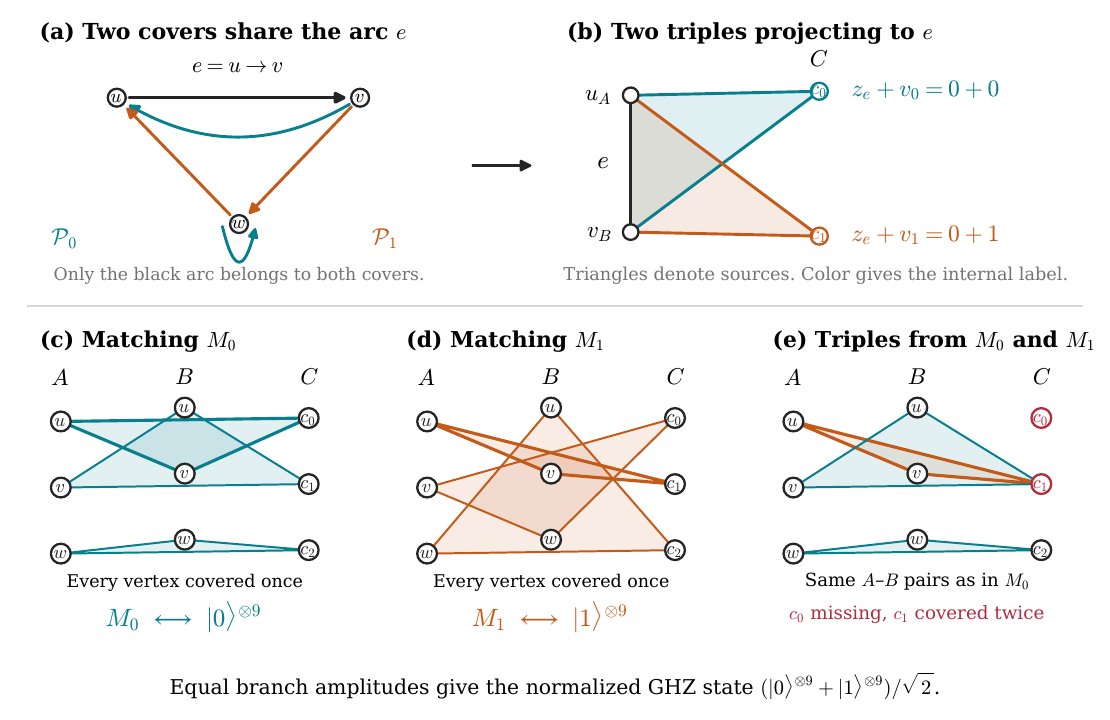}
\caption{A choice of triples that covers the first two parts can miss a path in the third. (a) The two cycle covers share the arc $u\to v$. (b) The triples that project to this arc use $c_0$ in $M_0$ and $c_1$ in $M_1$. Each triangle represents one source. Its color gives the internal label of the 3 photons, one in each path of the triple. In panels (c)--(e), each of $u,v,w$ denotes one path in $A$ and a different path in $B$. (c,d) When precisely the sources in $M_i$ emit, every path receives one photon and the output label state is $|i\rangle^{\otimes9}$, for $i=0,1$. (e) Switching only the triple that projects to $u\to v$ covers $c_1$ twice and leaves $c_0$ uncovered.}
\label{fig:comb}
\end{figure}

Identify each vertex of $C$ with its tag vector. A perfect matching uses every tag once, so its tags sum to $\sum_{c\in C}c$. We choose the translations so that a mixed choice of triples covering the first two parts exactly once has a different tag sum.

\begin{samepage}
\begin{lemma}\label{lem:exposed-translations}
Let $D$ be a digraph on $m$ vertices with loops allowed and no parallel arcs, and let $\mathcal P_1,\ldots,\mathcal P_q$ be all its distinct cycle covers. Fix a vector $\mathbf z_e$ for each arc and translations $\mathbf v_1,\ldots,\mathbf v_q$ in the same real vector space. Suppose that for each $i$ some linear functional takes a strictly smaller value at $\mathbf v_i$ than at every $\mathbf v_j$ with $j\ne i$. Let $C$ be an $m$-element subset of this space such that
\begin{equation}
 \{\mathbf z_e+\mathbf v_i\colon e\in\mathcal P_i\}=C\qquad(i\in[q]).
 \label{eq:con-selector-tags}
\end{equation}
Take two copies $A,B$ of $V(D)$ and a third part $C$. For each $i\in[q]$ and each arc $e=u\to v$ of $\mathcal P_i$, include the triple $(u_A,v_B,\mathbf z_e+\mathbf v_i)$. The resulting simple tripartite hypergraph is $q$-rigid.
\end{lemma}
\end{samepage}
\begin{proof}
By \cref{eq:con-selector-tags}, the $m$ triples from each cover form a perfect matching. Triples above the same arc from different covers have distinct tags, so these matchings are edge-disjoint. Any perfect matching projects to some $\mathcal P_i$. Both this matching and the prescribed matching above $\mathcal P_i$ use every element of $C$ once. For $e\in\mathcal P_i$, let $j(e)$ be the index of the prescribed matching that contains the chosen triple above $e$. Subtracting the two tag sums gives $\sum_{e\in\mathcal P_i}(\mathbf v_{j(e)}-\mathbf v_i)=\mathbf0$, since the anchors cancel. Apply a linear functional whose unique minimum on $\{\mathbf v_1,\ldots,\mathbf v_q\}$ occurs at $\mathbf v_i$. Each scalar summand is nonnegative and vanishes only if $j(e)=i$. Hence the matching consists of the triples from $\mathcal P_i$.
\end{proof}

For $t^2$ prescribed covers, put $\gamma_j\coloneqq\binom j2$ and take $\mathbf v_{ab}\coloneqq(a,b,\gamma_a+\gamma_b)$ for $0\le a,b<t$. The functional $\lambda_{ab}(x,y,z)\coloneqq2z-(2a-1)x-(2b-1)y$ satisfies
\begin{equation}
 \lambda_{ab}(\mathbf v_{a'b'}-\mathbf v_{ab})
 =(a'-a)^2+(b'-b)^2.
 \label{eq:main-exposure}
\end{equation}
Once each prescribed cover uses every tag in $C$ exactly once, the sum of squares in \cref{eq:main-exposure} rules out any mixed perfect matching. It remains to choose anchors satisfying \cref{eq:con-selector-tags} for every cover.

For the construction with $t^2$ prescribed matchings, we use the projected digraph shown in \cref{fig:doublecomb-routes}, which we call the \emph{double comb}. For the route indexed by $(a,b)$, follow the upper chain of row $a$, leave at exit $b$, traverse the vertices of $K_{ab}$ in order, and return along the lower chain to the common vertex $s$. Every directed cycle of length at least 2 follows one of these routes through $s$. Since $s$ has no loop, a cycle cover uses exactly one such route. Loops cover the vertices outside the route. Each of the $t$ rows has $t$ exits, so there are $t^2$ projected perfect matchings. With the tag assignment in Appendix~\ref{app:construction}, each cycle cover lifts to a unique perfect matching and hence gives one GHZ branch.

\begin{figure}[htbp]
\centering
\includegraphics[width=\linewidth]{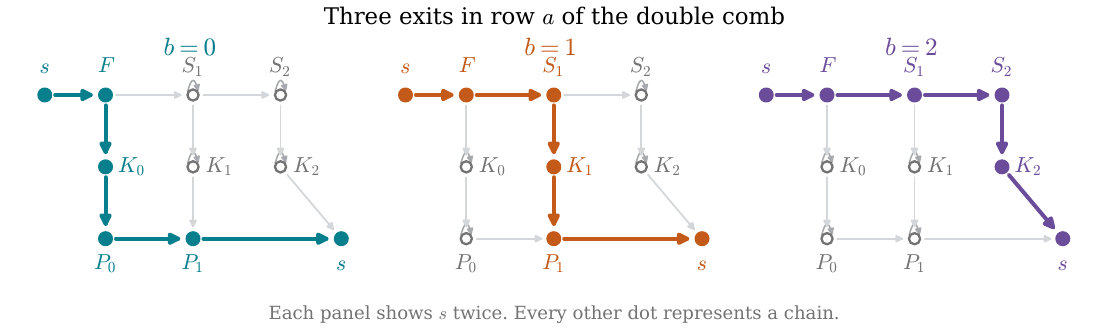}
\caption{Three exits $b=0,1,2$ in one row of the double comb. The route to exit $b$ traverses the common chain $F$ and its private block $K_b$. It visits $S_j$ for $j\le b$ and $P_\ell$ for $\ell\ge b$. A gray loop at a contracted chain represents one loop at each of its unused vertices.}
\label{fig:doublecomb-routes}
\end{figure}

Within a row, a shared arc occurs on the routes for all exits, for a prefix of the exits, or for a suffix. We take the tags from a box of lattice points with selected points removed. For a loop at $x$ and a fixed row $i$, the exits $b$ for which $x+\mathbf v_{ib}$ lies in a given removed rectangle form an interval. Counting these intervals shows that enough anchors remain for the shared arcs. In each cover we assign the remaining tags bijectively to the arcs of its private path. Appendix~\ref{app:construction} gives the details and the bound of $O(t^7)$ vertices per part.

For $m\ge2$, a cyclic extension gives $g_{r+1}(m)\ge2g_r(m)$. To see this, take a balanced $r$-partite rigid $r$-graph and index one part as $x_0,\ldots,x_{m-1}$. Add a part $y_0,\ldots,y_{m-1}$. Extend each edge through $x_i$ in two ways, by $y_i$ and by $y_{i+1}$, with indices modulo $m$. Fix a perfect matching in the extended hypergraph. It projects to an old perfect matching. If its edge through $x_i$ uses $y_{i+1}$, its edge through $x_{i+1}$ must use $y_{i+2}$. Going around the cycle shows that either the matching pairs $x_i$ with $y_i$ for every $i$, or it pairs $x_i$ with $y_{i+1}$ for every $i$. Each old matching has precisely these two extensions, and their edge sets are disjoint. The part size stays fixed as the total path count increases from $rm$ to $(r+1)m$.

\subsection{A product on two common parts}

Call a set $U$ of projection vertices an \emph{active support} if the loop at each vertex outside $U$ is its only incident arc. The factors can share their first two parts when the sum of their active support sizes is at most the common part size $m$.

\begin{theorem}\label{thm:support-product}
Let $k,m,q_1,\ldots,q_k$ be positive integers. For $f\in[k]$, let $H_f$ be a simple $q_f$-rigid tripartite $3$-graph with parts $A_f,B_f,C_f$ of size $m$. Identify $A_f$ and $B_f$ with two copies of a set $W_f$. Suppose there is a set $U_f\subseteq W_f$ such that every edge $(u_{A_f},v_{B_f},c)$ satisfies either $u,v\in U_f$ or $u=v\in W_f\setminus U_f$. If $\sum_{f\in[k]}|U_f|\le m$, there is a balanced $(k+2)$-partite hypergraph of uniformity $k+2$ and part size $m$ that is $\prod_{f\in[k]}q_f$-rigid.
\end{theorem}

Place the sets $U_f$ disjointly in a common $m$-element set and use two copies of that set for the first two parts. Each factor keeps its own tag part $C_f$. With the edges defined in Appendix~\ref{app:construction}, choosing one perfect matching in each factor gives a perfect matching of the product, and every perfect matching arises from exactly one such choice.

Write $k=r-2$ and $T=t-1$, where $t$ is the number of exits in each row. We use a lattice box with two side lengths of order $kT^2$ and a third of order $kT^3$. The resulting tripartite factor has $(T+1)^2$ matchings. Its active support $U$ consists of the common vertex $s$ and the vertices whose loops are omitted from at least one prescribed cover. Since $k|U|\le m$, we can place $k$ copies on the two common parts. Their product has $(T+1)^{2k}$ matchings on parts of size $O(k^3T^7)$, giving the exponent $2k/7$. At smaller part sizes a planar construction uses one row and satisfies the same support condition. The two support estimates are proved in \cref{app:rect-support,app:planar}.

\section{Upper bounds and sharp small cases}\label{sec:finite}

Since distinct perfect matchings in a rigid hypergraph are edge-disjoint, an exchange that produces a different perfect matching must replace every edge of the original one. We use this restriction to bound the number of perfect matchings. On 9 paths we determine the maximum local dimension both for unrestricted triplet sources and for sources that feed one path in each of three fixed groups of size 3.

\subsection{An upper bound from partial exchanges}\label{sec:harmonic}

For bipartite graphs, Br{\`e}gman's permanent inequality~\cite{Bregman1973} bounds the number of perfect matchings in terms of the degrees on one side. Alon and Friedland~\cite{AlonFriedland2008} obtained a degree bound for general graphs, and Taranenko~\cite{Taranenko2017} bounded the number of perfect matchings in uniform hypergraphs using multidimensional permanents. Set $\beta_1\coloneqq1$. For $m\ge2$, let $\beta_m$ be the smaller of $m/2$ and the harmonic sum $\sum_{j=1}^m 1/j$.

For an $r$-graph $H$, its $(r-1)$-shadow $\partial_{r-1}H$ consists of the $(r-1)$-sets contained in an edge. For $S\in\partial_{r-1}H$, its codegree $d_H(S)$ is the number of edges containing $S$.

\begin{theorem}\label{thm:dimension-upper}
Let $r\ge2$ and $m\ge1$ be integers, and let $H$ be a $d$-rigid $r$-graph on $rm$ vertices. Then
\begin{equation}
 d\beta_m \le\sum_{S\in\partial_{r-1}H}
 \left[1-\left(1-\frac1r\right)^{d_H(S)}\right]
 \le |\partial_{r-1}H|\le\binom{rm}{r-1}.
 \label{eq:harmonic-shadow}
\end{equation}
If $H$ is balanced $r$-partite with parts $V_1,\ldots,V_r$, then $d\beta_m\le\bigl|\{e\setminus V_1\colon e\in H\}\bigr|\le m^{r-1}$.
Consequently,
\begin{equation}
 f_r(rm)\le\left\lfloor\frac{1}{\beta_m}\binom{rm}{r-1}\right\rfloor,\qquad
 g_r(m)\le\left\lfloor\frac{m^{r-1}}{\beta_m}\right\rfloor.
 \label{eq:harmonic-extremal}
\end{equation}
\end{theorem}

\begin{proof}
For $m=1$, the full vertex set is the only possible edge, so $d$ is 0 or 1. The sum in \cref{eq:harmonic-shadow} equals $d$, as does the number of distinct projections $e\setminus V_1$ in the partite case.

Suppose $m\ge2$. For each perfect matching $M$ of $H$, construct a digraph as follows. For $e\in M$ and $v\in e$, draw an arc $v\to w$ if $w\notin e$ and $(e\setminus\{v\})\cup\{w\}$ is an edge of $H$. Choose a set $Q$ consisting of one representative from each edge of $M$, and restrict this digraph to $Q$. Write $d_Q^+(v)$ for the outdegree of $v$ in the restriction.

A directed cycle of length less than $m$ would let us replace each representative on the cycle by its successor, producing a different perfect matching with at least one edge of $M$ unchanged. Hence the restricted digraph is acyclic or consists of a directed $m$-cycle. Any additional arc between vertices of such a cycle would create a shorter directed cycle. In the acyclic case a topological ordering $v_1,\ldots,v_m$ gives $d_Q^+(v_i)\le m-i$, so the sum below is at least $\sum_{j=1}^m 1/j$. On the directed cycle every outdegree is 1 and the sum is $m/2$. In either case,
\begin{equation}
 \sum_{v\in Q}\frac1{1+d_Q^+(v)}\ge\beta_m.
 \label{eq:harmonic-acyclic}
\end{equation}

Choose the representatives independently and uniformly within their matching edges. Fix $e\in M$ and $v\in e$, and put $c=d_H(e\setminus\{v\})$. For each other edge $F\in M$, let $k_F$ count the outneighbors of $v$ in $F$. Conditional on $v\in Q$, its outdegree $d_Q^+(v)$ is a sum of independent Bernoulli variables with probabilities $k_F/r$, where $0\le k_F\le r$ and $\sum_F k_F=c-1$. Hence
\begin{align*}
 \mathbb E\!\left[\frac1{1+d_Q^+(v)}\,\middle|\,v\in Q\right]
 &=\int_0^1\prod_F\left(1-\frac{k_F}{r}(1-z)\right)\,\mathrm dz\\
 &\le\int_0^1\left(1-\frac{1-z}{r}\right)^{c-1}\,\mathrm dz
 =\frac r c\left[1-\left(1-\frac1r\right)^c\right].
\end{align*}
The inequality uses $1-ku\le(1-u)^k$ for integers $0\le k\le r$ and $0\le u\le1/r$. Since $\mathbb P[v\in Q]=1/r$, taking expectations in \cref{eq:harmonic-acyclic} yields
\[
 \beta_m\le\sum_{e\in M}\sum_{S\in\binom e{r-1}}
 \frac{1-\bigl(1-\tfrac1r\bigr)^{d_H(S)}}{d_H(S)}.
\]
Now sum over the $d$ perfect matchings, which partition $H$. For each $(r-1)$-set $S$ in the shadow, the sum contains $d_H(S)$ identical terms, each with a factor $1/d_H(S)$. This proves \cref{eq:harmonic-shadow}.

For the partite bound, take $Q=V_1$. For $e\in M$, let $v$ be its unique vertex in $V_1$. All vertices completing $e\setminus\{v\}$ lie in $V_1$, so $1+d_Q^+(v)=d_H(e\setminus\{v\})$. Summing \cref{eq:harmonic-acyclic} over the $d$ matchings counts each distinct set of the form $e\setminus V_1$, $e\in H$, with total weight 1. There are at most $m^{r-1}$ such sets. Taking the maximum over the respective classes gives \cref{eq:harmonic-extremal}.
\end{proof}

For $m\ge5$ we have $\beta_m=\sum_{j=1}^m 1/j=\log m+O(1)$. For fixed $r$ and $n=rm\to\infty$, \cref{eq:harmonic-extremal} gives $f_r(n)\le(1/(r-1)!+o(1))\,n^{r-1}/\log n$ and $g_r(m)\le(1+o(1))m^{r-1}/\log m$. For triples this gives $f_3(n)=o(n^2)$, answering a question of \texttt{discrete\_things}~\cite{DiscreteThings2024}. For $f_3(n)$, the lower bound from \cref{thm:growth} is $c_3n^{2/7}$ and the upper bound is $(1/2+o(1))n^2/\log n$.

At $m=2,3,4$, the values of $\beta_m$ are $1,3/2,2$, respectively. In particular, $g_r(3)\le2\cdot3^{r-2}$ and $g_3(4)\le8$. The six-matching construction in \cref{finite:thm:six} applies at every part size $m\ge3$. Thus on 12 paths with 4 paths in each of 3 groups we have $6\le g_3(4)\le8$.

\subsection{Nine paths}

\begin{theorem}\label{thm:finite-exact}
For unrestricted triplet sources on 9 paths in the model of \cref{sec:model}, the maximum local GHZ dimension $f_3(9)$ is 13. If every triple meets each of three fixed groups of size 3, the maximum $g_3(3)$ is 6. In this case the $6$-rigid hypergraph attaining this maximum is unique up to relabeling within each group.
\end{theorem}

Gu, Chen, and Krenn~\cite[Sec.~III\,B]{GuChenKrenn2020} reported that dimension 13 is attainable on 9 paths with triplet sources. In 2024, \texttt{discrete\_things}~\cite{DiscreteThings2024} listed 13 perfect matchings on 9 vertices whose union has no others. We construct a hypergraph with 13 perfect matchings on the affine plane $\mathbb F_3^2$, show that it is not isomorphic to the 2024 example, and prove $f_3(9)\le13$.

For unrestricted triplet sources, dimension 1 requires only 3 paths. The minimum is 6 paths for dimensions $2$ through $10$ and 9 paths for dimensions $11$ through $13$. Indeed, $f_3(3)=1$, $f_3(6)=10$, and deleting whole matching classes realizes any smaller positive dimension on the same paths. The 6-path networks use $2d$ source configurations, and the 9-path networks use $3d$.

\subsubsection*{Three fixed groups of paths}

\Cref{thm:dimension-upper} gives $g_3(3)\le 3^2/\beta_3=6$. To attain this bound, take three disjoint copies of $\mathbb F_3$ as the vertex parts. Let $H$ consist of the triples $(a,b,c)$ with $a+b+c\in\{0,2\}$. For $h\in\{0,2\}$ and $s\in\mathbb F_3$, set
\begin{equation}
 M_{h,s}\coloneqq\{(a,a+s,a+h-s)\colon a\in\mathbb F_3\}.
 \label{eq:finite-six-matchings}
\end{equation}
These 6 perfect matchings partition the 18 edges of $H$, since an edge determines $h=a+b+c$ and $s=b-a$.

It remains to check that $H$ has no other perfect matching. The 3 edge sums of any perfect matching add to zero in $\mathbb F_3$, since the matching covers every vertex. Each sum lies in $\{0,2\}$, so all 3 have the same value $h$. Write the edges of the matching as $(a,b(a),c(a))$, where $b$ and $c$ are permutations of $\mathbb F_3$. Every permutation of $\mathbb F_3$ has the form $a\mapsto\varepsilon a+s$ with $\varepsilon\in\{1,-1\}$. Write $b$ in this form. If $\varepsilon=-1$, the equation $a+b(a)+c(a)=h$ makes $c(a)=h-s$ constant, which is impossible. Thus $\varepsilon=1$, and the matching is one of those in \cref{eq:finite-six-matchings}. This proves $g_3(3)=6$.

Now let $H$ be a balanced tripartite $6$-rigid $3$-graph with 3 vertices per part. Take $Q=V_1$ in the proof of \cref{thm:dimension-upper}. Since $6\beta_3=9$, all 9 pairs in $V_2\times V_3$ occur, and each of the 6 matchings attains equality in \cref{eq:harmonic-acyclic}. An acyclic exchange digraph would contribute at least $1+1/2+1/3=11/6>3/2$, so each digraph is a directed triangle. Every vertex in each exchange digraph has outdegree 1. Hence every pair in $V_2\times V_3$ has exactly 2 completions in $V_1$. The same argument applies with any part as $Q$. Thus the nine missing triples contain every cross-part pair exactly once and specify a Latin square of order 3. After relabeling, the first row and column read $0,1,2$. The middle entry must be 2, since a 0 there would put a second 2 in the last column. This forces the unique reduced square of order 3, as recorded by McKay and Wanless~\cite{McKayWanless2005}. The other 18 triples of $V_1\times V_2\times V_3$ form the unique extremal hypergraph up to relabeling within parts.

\subsubsection*{An algebraic construction with 13 matchings}

We now allow a source to meet any 3 of the 9 paths. Identify the vertex set with $\mathbb F_3^2$. For $\boldsymbol\sigma\in\mathbb F_3^2$, define
\[
 H_{\boldsymbol\sigma}\coloneqq\left\{\{\mathbf x,\mathbf y,\mathbf z\}\in\binom{\mathbb F_3^2}{3}\colon \mathbf x+\mathbf y+\mathbf z=\boldsymbol\sigma\right\}.
\]
Take $H\coloneqq\bigcup_{\boldsymbol\sigma\in\{0,1\}^2}H_{\boldsymbol\sigma}$. We first determine the perfect matchings within each sum class and then show that every perfect matching of $H$ lies in one class.

The triples in $H_{\mathbf0}$ are the 12 affine lines of $\mathbb F_3^2$. Two such lines are disjoint precisely when they are parallel. Thus the 4 parallel classes are exactly the perfect matchings of $H_{\mathbf0}$, and they partition its edges.

Fix $\boldsymbol\sigma\ne\mathbf0$. Choose $\mathbf u\notin\langle\boldsymbol\sigma\rangle$ and put $\mathbf v\coloneqq\boldsymbol\sigma-\mathbf u$. The vectors $\mathbf u,\mathbf v$ are linearly independent. Set $T\coloneqq\{\mathbf0,\mathbf u,\mathbf v\}$. Its 9 translates are distinct triples with sum $\boldsymbol\sigma$. Indeed, invariance under a nonzero translation would make $T$ an affine line, contrary to linear independence. There are $\binom93-12=72$ triples with nonzero sum, and invertible linear maps act transitively on the 8 nonzero sums. Consequently $H_{\boldsymbol\sigma}$ has exactly 9 edges, all of which are translates of $T$.

The difference set satisfies
\[
 (T-T)\setminus\{\mathbf0\}
 =\{\pm\mathbf u,\pm\mathbf v,\pm(\mathbf u-\mathbf v)\}
 =\mathbb F_3^2\setminus\{\mathbf0,\boldsymbol\sigma,-\boldsymbol\sigma\}.
\]
It follows that two distinct translates $T+\mathbf x$ and $T+\mathbf y$ are disjoint exactly when $\mathbf x-\mathbf y=\pm\boldsymbol\sigma$. A family of 3 translates forms a perfect matching precisely when its translation vectors constitute a coset of $\langle\boldsymbol\sigma\rangle$. Thus $H_{\boldsymbol\sigma}$ has 3 perfect matchings, partitioning its 9 edges.

Finally, let $M$ be any perfect matching of $H$, and let $\boldsymbol\sigma_1,\boldsymbol\sigma_2,\boldsymbol\sigma_3\in\{0,1\}^2$ be its edge sums. Their total is the sum of all vertices, which is $\mathbf0$. In each coordinate, 3 elements of $\{0,1\}$ sum to zero in $\mathbb F_3$ only when they are all equal. Hence $\boldsymbol\sigma_1=\boldsymbol\sigma_2=\boldsymbol\sigma_3$. Every perfect matching lies in one sum class. The construction therefore has exactly $4+3+3+3=13$ perfect matchings, and these matchings partition its 39 edges.

Number the path $(x,y)$ by $3x+y$, using the integer representatives $0,1,2$ for $x,y\in\mathbb F_3$. Assign internal label $i$ to the three sources in matching $M_i$ of \cref{tab:finite-thirteen}. These 39 triplet-source configurations give the 13 GHZ branches in the model of \cref{sec:model}.

To compare this hypergraph with the one listed by \texttt{discrete\_things}~\cite{DiscreteThings2024}, join each pair of vertices of codegree 4. In each case the resulting graph consists of 3 disjoint triangles. In \cref{tab:finite-thirteen}, the vertex set of each of the 3 triangles is a hyperedge. In the 2024 example, this holds for only 2 of the triangles. Thus the hypergraphs are nonisomorphic.

\begin{table}[htbp]
\centering
\caption{The 13 perfect matchings of the algebraic construction. A string such as $012$ denotes the triple $\{0,1,2\}$. The 13 indexed matchings are pairwise edge-disjoint.}
\label{tab:finite-thirteen}
\begin{tabular}{c@{\qquad}ccc@{\qquad\qquad}c@{\qquad}ccc}
\toprule
Index&\multicolumn{3}{c}{Source configurations}&Index&\multicolumn{3}{c}{Source configurations}\\
\midrule
0&012&345&678&7&037&148&256\\
1&036&147&258&8&046&157&238\\
2&048&156&237&9&058&136&247\\
3&057&138&246&10&013&268&457\\
4&015&267&348&11&025&178&346\\
5&024&168&357&12&067&124&358\\
6&078&123&456&&&&\\
\bottomrule
\end{tabular}
\end{table}

\Cref{fig:sourceevents} shows two prescribed emission patterns and a mixed pattern with the same total photon number. Of these three, only the prescribed patterns place one photon in every path. When all 13 accepted branches have the same amplitude, the normalized postselected state is $|\mathrm{GHZ}_{9,13}\rangle$.

\begin{figure}[htbp]
\centering
\includegraphics[width=\linewidth]{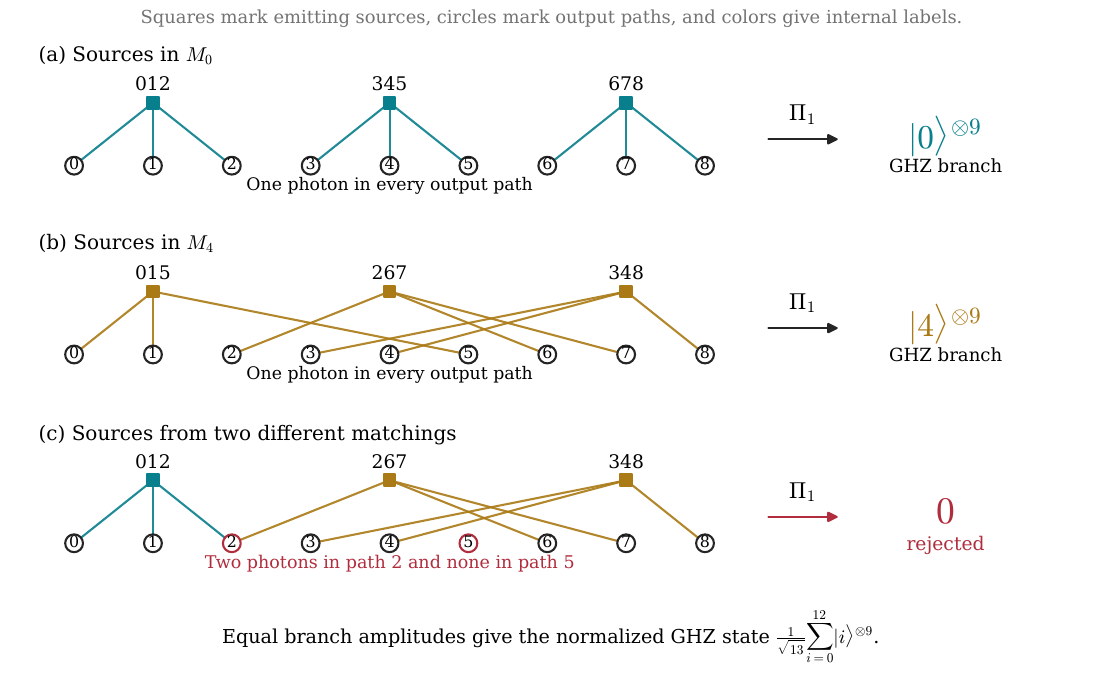}
\caption{Source configurations and photon-number projection in the 13-matching construction. Squares denote emitting triplet sources, circles denote output paths, and colors specify the internal labels of the emitted photons. (a,b) After projection, the emission patterns $M_0$ and $M_4$ from \cref{tab:finite-thirteen} contribute the basis states $|0\rangle^{\otimes9}$ and $|4\rangle^{\otimes9}$, respectively. (c) The mixed event sends 2 photons to path $2$ and none to path $5$, so $\Pi_1$ sends the corresponding state to zero. The formula below the panels gives the equal-weight superposition of all 13 accepted branches.}
\label{fig:sourceevents}
\end{figure}

\subsubsection*{The unrestricted upper bound}

Suppose a rigid $3$-graph on 9 vertices has at least 14 perfect matchings. The union of any 14 is a $14$-rigid hypergraph $H$ with 42 edges and degree 14 at every vertex. Appendix~\ref{app:finite} shows that every pair has codegree at most 4. The average pair codegree is $3\cdot42/36=7/2$, so some pair $ab$ has codegree 4. The 18 triples avoiding $a$ and $b$ are then determined up to relabeling. Regularity requires 10 edges meeting $\{a,b\}$ at $a$ alone and 10 at $b$ alone, whereas this configuration forces one of these counts to be at most 9. Thus $f_3(9)\le13$.

The $13$-rigid triple system in \cref{finite:prop:thirteen18} gives $f_3(18)\ge13$. Grafting a balanced tripartite $d$-rigid hypergraph of part size $q$ onto a $d$-rigid $3$-graph on $3m$ vertices gives a $d$-rigid $3$-graph on $3(m+q-1)$ vertices (\cref{finite:sec:grafting}). In \cref{finite:thm:ten-all} we use grafting to prove $f_3(3m)\ge10$ for every $m\ge2$.

\section{Source configurations and optimal emission probabilities}\label{sec:resources}

Fix a $d$-rigid network with $m=n/r$ sources in each prescribed perfect matching. An accepted event occurs when precisely the sources of one matching emit. In the independent binary-source model, $m$ and $d$ determine the maximum success probability when the accepted amplitudes all have the same magnitude. A phase correction on one output path then gives the state in \cref{eq:GHZ-definition}.

\subsection{Counting paths, sources, and modes}

In a $d$-rigid $r$-graph on $n=rm$ vertices, each edge specifies one $r$-photon source configuration. The network uses $dm$ configurations in all and $nd$ orthogonal modes, one for each choice of a path and a matching label. A device can supply several configurations. In the frequency-comb proposal of Zhu et al.~\cite{ZhuEtAl2020}, one microring supplies several frequency pairs for a polarization GHZ state. Gu and Krenn~\cite{GuKrennCombs2020} describe the perfect matchings in this scheme. Arlt, Ruiz-Gonzalez, and Krenn~\cite{ArltRuizGonzalezKrenn2025} proposed networks of pair sources that emulate a 4-, 6-, or 8-photon emitter.

For fixed $r\ge3$, the networks in \cref{cor:prescribed} realize every $d\ge2$ with numbers of paths and source configurations bounded by polynomials in $d$. Conversely, every $d$-rigid network of $r$-photon sources satisfies $d=O_r(n^{r-1}/\log n)$ by \cref{thm:dimension-upper}, so
\[
 n=\Omega_r\!\left((d\log d)^{1/(r-1)}\right).
\]
Indeed, $d\le\binom{n-1}{r-1}\le n^{r-1}$ gives $\log n\ge(\log d)/(r-1)$. Substituting into $n^{r-1}=\Omega_r(d\log n)$ gives the claimed lower bound on the number of paths.

\subsection{Equal-weight branches in the binary-source model}

Suppose first that the sources in \cref{eq:normalized-source-state} have the same emission probability $p$ and that their relative phases $\phi_e$ are equal. Each prescribed matching uses $m$ of the $dm$ sources. The matchings give mutually orthogonal output states, so their probabilities add to give the success probability
\begin{equation}
 P_{m,d}(p)=d\,p^m(1-p)^{m(d-1)}.
 \label{eq:binary-success}
\end{equation}
For $d\ge2$, differentiating its logarithm gives $m/p-m(d-1)/(1-p)$. This derivative vanishes only at $p=1/d$, is positive below this point, and is negative above it. Thus
\begin{equation}
 P_{m,d}^{\max}=d^{1-m}\left(1-\frac1d\right)^{m(d-1)}.
 \label{eq:binary-success-optimal}
\end{equation}
At the optimum, the expected number of emitting sources is $dm/d=m$, so the expected total photon number is $rm=n$.

Allowing the sources to have different emission probabilities gives the same maximum when all branches must have equal weight.

\begin{proposition}\label{prop:balanced-pump}
Let $H$ be a $d$-rigid $r$-graph on $n=rm$ vertices, with $d\ge2$, and let its sources satisfy \cref{eq:normalized-source-state}. Among choices $0<p_e<1$ for which all accepted branches have equal probability, the maximum success probability is given by \cref{eq:binary-success-optimal}. This maximum is attained exactly when $p_e=1/d$ for every $e\in H$.
\end{proposition}
\begin{proof}
Set $x_e\coloneqq p_e/(1-p_e)$. By \cref{eq:matching-amplitude}, the probability that precisely the sources in $M_i$ emit is
\[
 |B_{M_i}|^2=\frac{\prod_{e\in M_i}x_e}{\prod_{e\in H}(1+x_e)}.
\]
The branch probabilities have a common denominator, so equal probabilities imply that the geometric mean $z=(\prod_{e\in M_i}x_e)^{1/m}$ is independent of $i$. Jensen's inequality for the strictly convex function $y\mapsto\log(1+e^y)$, applied to $\log x_e$ for $e\in M_i$, gives $\prod_{e\in M_i}(1+x_e)\ge(1+z)^m$. Multiplying over the $d$ matchings yields $\prod_{e\in H}(1+x_e)\ge(1+z)^{dm}$. Each numerator equals $z^m$, so by \cref{eq:binary-success,eq:binary-success-optimal} the success probability $P=dz^m/\prod_{e\in H}(1+x_e)$ satisfies
\[
 P\le\frac{d\,z^m}{(1+z)^{dm}}
   =P_{m,d}\!\left(\frac{z}{1+z}\right)\le P_{m,d}^{\max}.
\]
Equality in the first inequality holds exactly when $x_e=z$ for every edge. Equality in the second inequality holds exactly when $z/(1+z)=1/d$. Thus equality throughout is equivalent to $p_e=1/d$ for every $e\in H$.
\end{proof}

For fixed $r,n,d$, every $d$-rigid network uses $dn/r$ source configurations and, by \cref{prop:balanced-pump}, has the same optimal success probability in the independent binary-source model when the branches are equally likely. \Cref{tab:twelve-networks} compares three networks with $d=12$. Deleting one matching from the 13-matching example gives the 9-path network. The 39- and 117-path networks are balanced tripartite. The 117-path example is also \emph{linear}, meaning that any two distinct source supports meet in at most one path.

\begin{table}[htbp]
\centering
\caption{These three triplet-source networks each attain local dimension 12. On $n$ paths, each matching uses $m=n/3$ of the $12m$ source configurations. The last column gives the maximum success probability per trial under ideal transmission and detection, subject to equal branch probabilities, as in \cref{eq:binary-success-optimal}.}\label{tab:twelve-networks}
\begin{tabular}{@{}lrrr@{}}
\toprule
Restrictions on source supports & Paths $n$ & Source configurations & $P_{m,12}^{\max}$\\
\midrule
Unrestricted        & 9   & 36  & $3.93\times10^{-4}$\\
Balanced tripartite & 39  & 156 & $4.43\times10^{-19}$\\
Linear balanced tripartite & 117 & 468 & $6.02\times10^{-58}$\\
\bottomrule
\end{tabular}
\end{table}

In each dimension-13 network, a successful trial requires precisely the sources in one prescribed matching to emit. Such a matching uses 3 of the 39 sources on 9 paths and 6 of the 78 on 18 paths. The corresponding maximum success probabilities are $P_{3,13}^{\max}\simeq3.32\times10^{-4}$ and $P_{6,13}^{\max}\simeq8.46\times10^{-9}$ (\cref{prop:balanced-pump}).

For fixed $r$ and $\alpha>0$, suppose $d=\Theta(n^\alpha)$. Substituting $m=n/r$ into \cref{eq:binary-success-optimal} gives
\[
 \log P_{m,d}^{\max} =(1-m)\log d+m(d-1)\log\!\left(1-\frac1d\right)
 =-\frac{\alpha}{r}\,n\log n+O(n).
\]
From the construction in \cref{thm:growth}, retain $d=\Theta(n^{2(r-2)/7})$ perfect matchings. In this family, $\log P_{m,d}^{\max}=-\frac{2(r-2)}{7r}n\log n+O(n)$. For the triplet networks in this family, the coefficient of $-n\log n$ is $2/21$. These networks use $\Theta(n^{1+2(r-2)/7})$ source configurations for fixed $r$.

\section*{Acknowledgments}

The second author thanks Rishikesh Gajjala, whose 2024 question about the number of perfect matchings in rigid $3$-uniform hypergraphs motivated this work.

H.L. was supported by the National Natural Science Foundation of China (12501487), the China Scholarship Council, and the Institute for Basic Science (IBS-R029-C4). X.L. was supported by the Excellent Young Talents Program (Overseas) of the National Natural Science Foundation of China.

\section*{Data and code availability}

The \href{https://github.com/xliu2022/xliu2022.github.io/tree/main/GHZ_supplementary_data}{data repository} and \ghzdataarchive{} contain the source lists, network export program, and table of lower bounds for $1\le m\le1000$. The \ghzdatafile{README.txt}{instructions for exporting the networks} give the commands and describe the source-list format. The \ghzdatafile{certificates.md}{certificate index} lists finite matching families, transfer matrices, and the residue representatives for the linear thresholds in \cref{finite:thm:linear-thresholds}. The code includes exact-arithmetic checks of the rectangle counts and support estimates in Appendix~\ref{app:construction} and two independent C++17 programs verifying the nine-path upper bound in Appendix~\ref{app:finite}.

\appendix

\section{Construction at every part size}\label{app:construction}

Put $k\coloneqq r-2$ and $u\coloneqq(m/(128k^3))^{2/7}$. To prove \cref{thm:growth}, we seek at least $u^k$ perfect matchings on $m$ vertices per part. A single matching suffices for $u\le1$, and cyclic extension gives the bound for $1<u\le2$. For $2<u<224^2$ we use a planar factor with one row of routes. For $u\ge224^2$ we use a rectangular factor in which the row and the exit can be chosen independently. Both factors have exactly $m$ vertices per part and an active support $U$ with $k|U|\le m$, so \cref{thm:support-product} applies to $k$ copies. Throughout this appendix $[a,b]$ denotes the integer interval from $a$ to $b$, empty if $a>b$, and $[q]_0=\{0,\ldots,q-1\}$.

\subsection{The product on two common parts}

\begin{proof}[Proof of \cref{thm:support-product}]
Since $\sum_f|U_f|\le m$, we can identify each $W_f$ with the same $m$-element set $\Omega$, choosing the bijections so that the images of the $U_f$ are pairwise disjoint. Use these labels in both of the first two parts of each factor, and keep the tag parts $C_1,\ldots,C_k$ disjoint. In the $i$th matching $M_{f,i}$ of $H_f$, where $1\le i\le q_f$, write the triple with first coordinate $u$ as $(u,p_{f,i}(u),c_{f,i}(u))$. Here $c_{f,i}\colon\Omega\to C_f$ is a bijection, and $p_{f,i}$ is a permutation of $\Omega$ that fixes every point outside $U_f$.

For a tuple $\mathbf i=(i_1,\ldots,i_k)$ of matching indices, let $p_{\mathbf i}$ agree with $p_{f,i_f}$ on each $U_f$ and fix all remaining points. This is a permutation, since each $p_{f,i_f}$ preserves $U_f$. Take two copies $A,B$ of $\Omega$ and form
\[
 M_{\mathbf i}\coloneqq\bigl\{
 \bigl(u_A,p_{\mathbf i}(u)_B,c_{1,i_1}(u),\ldots,c_{k,i_k}(u)\bigr)
 \colon u\in\Omega\bigr\}.
\]
Each $M_{\mathbf i}$ is a perfect matching because every coordinate map is a bijection. Let $H$ be their union.

Project an edge $(u_A,v_B,c_1,\ldots,c_k)$ to factor $f$ by taking $(u,v,c_f)$ if $u\in U_f$ and $(u,u,c_f)$ otherwise. An edge of $M_{\mathbf i}$ projects to the triple of $M_{f,i_f}$ at $u$. Thus an edge common to two prescribed global matchings would belong, after projection, to the corresponding matchings in every factor. Since the matchings in each factor are edge-disjoint, the two index tuples must agree.

Let $M$ be any perfect matching of $H$. Every edge whose first coordinate lies in $U_f$ also has its second coordinate in $U_f$. Hence the permutation from $A$ to $B$ induced by $M$ restricts to a permutation of $U_f$. Projecting to factor $f$ keeps this permutation on $U_f$ and uses the identity on its complement. The projection also covers $C_f$ once, so it is a perfect matching of $H_f$, say $M_{f,i_f}$. If an edge of $M$ comes from $M_{\mathbf j}$, its projection also belongs to $M_{f,j_f}$. Edge-disjointness gives $j_f=i_f$ for every $f$. Thus $M$ consists of the $m$ edges of $M_{\mathbf i}$.
\end{proof}

\subsection{A rectangular construction}\label{app:rectangular}

\begin{proposition}\label{prop:rectangular-factor}
Let $k\ge1$ and $m\ge128k^3\,224^7$ be integers, and put $T=\lfloor(m/(128k^3))^{1/7}\rfloor$. There is a balanced tripartite hypergraph of part size $m$ that is $(T+1)^2$-rigid and has an active support $U$ satisfying $k|U|\le m$.
\end{proposition}

Put $t=T+1$, $\gamma_b=\binom b2$, and $D=\gamma_T$. We use the translations $\mathbf v_{ib}=(i,b,\gamma_i+\gamma_b)$ of \cref{eq:main-exposure}, with $0\le i,b\le T$. Let
\[
 A=4kT^2,\qquad B=6kT^2,\qquad
 H=\left\lfloor\frac{m-1}{AB}\right\rfloor,\qquad \rho=m-1-ABH.
\]
Let $X$ consist of $[A]_0\times[B]_0\times[H]_0$ and any $\rho$ points in its next layer $[A]_0\times[B]_0\times\{H\}$. The two projection parts are copies of $X\cup\{s\}$, where $s$ is the hub in \cref{fig:doublecomb-routes}. The tag set $C$ will consist of $m$ points of the box $Y=[A+T]_0\times[B+T]_0\times[0,H+2D]$. We construct $C$ by removing points from $Y$. A loop at $u\in X$ receives the tag $u+\mathbf v_{ib}$ in the cycle cover with row $i$ and exit $b$ whenever that cover uses the loop.

\paragraph{Removed points and their owners.}
We remove rectangles from the planes with first coordinate $t,2t,\ldots$. We call these planes the primary levels. The table gives the ranges of the second and third coordinates for each type of rectangle.
\[
\begin{array}{c@{\qquad}ll}
 &\text{second-coordinate range}&\text{third-coordinate range}\\
 \text{horizontal}&[(2a+1)T,(2a+2)T-1]&[2D,H-1]\\
 \text{vertical}&[T,B-1]&[(2a+2)D,(2a+3)D-1]\\
 \text{full}&[T,B-1]&[2D,H-1].
\end{array}
\]
Here $0\le a<L_h=B/(2T)$ on a horizontal level and $0\le a<L_v=\lfloor(H-D)/(2D)\rfloor$ on a vertical level. We will also use at most one shorter rectangle of the same width as a full rectangle. All primary levels have first coordinate at most $A-1$, so the inverse translates of their rectangles lie in the full layers of $X$.

In row $i$, a point $x\in X$ is an \emph{owner} of a primary rectangle $\mathcal R$ if $x+\mathbf v_{ib}\in\mathcal R$ for some exit $0\le b\le T$. The first coordinate of any owner in row $i$ is congruent to $-i$ modulo $t$. The inverse translates of rectangles on different primary levels lie in disjoint intervals of first coordinates. On a given level the unused blocks separate the inverse translates of distinct rectangles. Thus each owner is assigned to a unique rectangle and row.

For an owner $x$ of $\mathcal R$ in row $i$, the exits $b$ for which $x+\mathbf v_{ib}\in\mathcal R$ form a nonempty interval. If this interval is a proper prefix $[0,\ell]$ or suffix $[j,T]$, leave the loop at $x$ out of precisely the cycle covers at those exits. In every other case, leave it out of every cycle cover in the row. If the interval is internal, say $[a,b']$, the point $x$ can serve as a \emph{residual anchor} for a shared arc on each of $[0,a-1]$ and $[b'+1,T]$. We count its uses separately for the two patterns. The net demand of a prefix or suffix is the number of owners assigned that pattern minus the number of residual anchors available for it.

\begin{lemma}\label{lem:rectangle-owners}
Fix a row $i$ and a primary rectangle $\mathcal R$ of width $W\ge T$ and height $J\ge D$. Group the owners by the interval of exits at which their translates lie in $\mathcal R$. For $0\le\ell<T$, each of the patterns $[0,\ell]$ and $[\ell+1,T]$ has net demand $J-D+(W-T+1)\ell$. The number of owners whose loops are omitted at every exit in row $i$ is $(W-T)(J-D)+(T-1)D$.
\end{lemma}
\begin{proof}
For a fixed $x$, the exits $b$ for which $x+\mathbf v_{ib}$ lies in $\mathcal R$ form the intersection of the intervals imposed by the second and third coordinates. If an owner's interval is $[0,\ell]$, either the horizontal interval ends at $\ell$ and the vertical one contains $[0,\ell]$, or the vertical interval ends at $\ell$ and the horizontal one contains $[0,\ell+1]$. For an internal interval starting at $\ell+1$, a suffix in one coordinate overlaps a prefix in the other away from both endpoints of $[0,T]$. Counting the owners in each case gives
\[
\begin{aligned}
 \#\{\text{owners with interval }[0,\ell]\}
   &=J-\gamma_\ell+\ell(W-\ell-1),\\
 \#\{\text{owners with internal interval starting at }\ell+1\}
   &=D-\gamma_{\ell+1}+\ell(T-\ell-1).
\end{aligned}
\]
Since $\gamma_{\ell+1}-\gamma_\ell=\ell$, their difference is $J-D+(W-T+1)\ell$. For each exit $b$, exactly $WJ$ points $x\in X$ satisfy $x+\mathbf v_{ib}\in\mathcal R$. Thus as many owners have intervals ending at $\ell$ as have intervals starting at $\ell+1$. From each side subtract both the number of internal intervals ending at $\ell$ and the number starting at $\ell+1$. The two sides are then the net demands for $[0,\ell]$ and $[\ell+1,T]$.

There are $(W-T)(J-D)$ owners with a full interval. The other owners whose loops are omitted at every exit have internal intervals, formed where a prefix in one coordinate overlaps a suffix in the other strictly inside $[0,T]$. Counting these overlaps gives $\sum_{b=0}^{T-1}\gamma_b+\sum_{a=1}^{T}(D-\gamma_a)=(T-1)D$ such owners.
\end{proof}

\paragraph{The shared arcs.}
In row $i$, put an owner on $F_i$ if its loop is omitted at every exit, on $P_{i,\ell}$ if it is omitted precisely at exits $[0,\ell]$, and on $S_{i,j}$ if it is omitted precisely at exits $[j,T]$. Fix an order on the vertices of each chain. Each choice of row $i$ and exit $b$ also has a private block $K_{ib}$ of vertices, which the route traverses in a fixed order. Use the double comb of \cref{fig:doublecomb-routes} with routes
\[
 s,F_i,S_{i,1},\ldots,S_{i,b},K_{ib},P_{i,b},\ldots,P_{i,T-1},s
 \qquad(0\le i,b\le T).
\]
The suffix list is empty at $b=0$ and the prefix list at $b=T$. Insert all arcs along these routes and a loop at every point of $X$. Removing $s$ and the loops leaves an acyclic digraph. A cycle through $s$ chooses one row, passes through one private block and returns to $s$. Hence each route, together with the loops off it, is a cycle cover and these are all the covers. The shared arcs entering or lying inside a full-row or suffix chain are used at exactly the exits assigned to that chain. For a prefix chain the same holds for its internal and outgoing shared arcs. Thus a chain with $a$ owners needs $a$ anchors for shared arcs with the chain's exit pattern. The arc entering $K_{ib}$ and the arc leaving it both belong to the private path.

Fix $i$. Each pattern has two sets of external anchors, listed below.
\[
\begin{array}{ll}
[0,T]&([-i,-1]\cup[A,A+T-i-1])\times[B]_0\times[H]_0,\\
     &[A]_0\times[B]_0\times([-\gamma_i,-1]\cup[H+1,H+D-\gamma_i]),\\[2pt]
[0,\ell]&[A]_0\times\{B+T-\ell-1\}\times[H]_0,\\
     &[A]_0\times[B]_0\times[H+1+2D-\gamma_i-\gamma_{\ell+1},H+2D-\gamma_i-\gamma_\ell],\\[2pt]
[j,T]&[A]_0\times\{-j\}\times[H]_0,\\
     &[A]_0\times[B]_0\times[-\gamma_i-\gamma_j,-\gamma_i-\gamma_{j-1}-1].
\end{array}
\]
For a fixed row, all the listed anchor sets are pairwise disjoint and lie outside $X$. The upper vertical intervals begin above $H$, so they also avoid the partial layer. At every exit in its pattern, each anchor translates to a point of $Y$. For an anchor in the first set listed for a pattern, either the first coordinate of its translate lies outside $[T,A-1]$ or its second coordinate lies outside $[T,B-1]$. For an anchor in the second set, its translated height lies below $2D$ or above $H$. We remove points only inside $[T,A-1]\times[T,B-1]\times[2D,H-1]$, so all these tags remain available.

\paragraph{Choosing the rectangles.}
Let $Q=B-2T+1$. For the moment allow real multiplicities $x,y,z$ for the horizontal, vertical, and full types, respectively. For the prefix $[0,\ell]$ and the suffix $[\ell+1,T]$, the external supply for each pattern is $AH+AB\ell$. The net demand is $x(H-3D+\ell)+yQ\ell+z(H-3D+Q\ell)$. For the full row the supply is $TBH+ABD$, and the demand is $(x+y+z)D(T-1)+z(B-2T)(H-3D)$. For proper prefixes and suffixes, both counts are affine in $\ell$. Equating their constant and linear coefficients gives two equations for $x,y,z$. Equating the full-row counts gives the third. The solution is
\begin{equation}\label{eq:rect-calibration}
\begin{gathered}
 S=\frac{AH}{H-3D},\qquad x=S-z,\qquad y=\frac{AB-x}{Q}-z,\\[2pt]
 z=\frac{TBH+ABD-D(T-1)\bigl(S+\tfrac{AB-S}{Q}\bigr)}
 {(B-2T)\bigl(H-3D-\tfrac{D(T-1)}{Q}\bigr)}.
\end{gathered}
\end{equation}
Take $n_h=\lfloor x\rfloor$ horizontal rectangles, $n_v=\lfloor y\rfloor-1$ vertical rectangles, and $n_f=\lfloor z\rfloor$ rectangles of the full type, filling each horizontal or vertical level before starting another. Set $\alpha=z-n_f$ and $L=\lfloor\alpha(H-2D)\rfloor$. If $L\ge D$, add the shorter rectangle $[T,B-1]\times[2D,2D+L-1]$ at a new level. The resulting number of primary levels is
\begin{equation}\label{eq:rect-level-count}
 U_0=\left\lceil\frac{n_h}{L_h}\right\rceil
     +\left\lceil\frac{n_v}{L_v}\right\rceil+n_f+\mathbf 1_{\{L\ge D\}}.
\end{equation}

These integer choices leave enough anchors for every pattern. If the shorter rectangle is present, it adds $L-D+Q\ell$ to the net demand for each of $[0,\ell]$ and $[\ell+1,T]$. This is at most $\alpha(H-3D+Q\ell)+(1-\alpha)Q\ell$. Omitting one vertical rectangle reduces the demand for each of these two patterns by $Q\ell$. The shorter rectangle's full-row demand minus $\alpha$ times that of a full rectangle is at most $-(1-\alpha)D(B-3T+1)\le0$. If the shorter rectangle is absent, rounding cannot increase demand. In \cref{app:rect-support} we verify $x>1$, $y>2$ and $z>T-1$, so all three types occur. Since $H>3D$ and $B>2T$, each proper prefix and suffix has an owner from a horizontal rectangle. A full rectangle supplies owners whose loops are omitted at every exit.

\paragraph{Completing the tag set.}
Fix a row and a primary rectangle of width $W$ and height $J$. Sum the net demands for the prefixes $[0,\ell]$ over $0\le\ell<T$ and add the full-row demand. The formulas in \cref{lem:rectangle-owners} and $\sum_{\ell=0}^{T-1}\ell=D$ give exactly its area $WJ$. At the real multiplicities in \cref{eq:rect-calibration}, each demand equals its external supply. The total primary area weighted by these multiplicities is therefore
\[
 \sum_{\ell=0}^{T-1}(AH+AB\ell)+(TBH+ABD)=T(A+B)H+2ABD.
\]
We must remove $|Y|-m$ points from $Y$. With real multiplicities, the weighted area of the primary rectangles is $R_*$ short of this total, where
\begin{equation}\label{eq:rect-residue}
 R_*=AB+(A+B)(2D+1)T+(H+2D+1)T^2-1-\rho.
\end{equation}
After rounding, let $R$ be the number of additional points to remove from $Y$ to leave exactly $m$ tags. We call these points the private holes. The difference $R-R_*$ is the decrease in primary area due to rounding. In particular $R\ge R_*>0$, since $\rho<AB$. Set $P_x=\lfloor(A-1)/t\rfloor$ and $P_y=\lfloor(B-1)/t\rfloor$. Take $R$ private holes from
\[
 \{at:U_0<a\le P_x\}\times\{at:1\le a\le P_y\}\times[2D,H-1].
\]
The next subsection verifies this set has at least $R$ points. Write $Z_K$ for the chosen set, and remove it together with the primary holes to obtain $|C|=m$. The private vertices of route $(i,b)$ form the set $K_{ib}=Z_K-\mathbf v_{ib}$ in $X$. The first coordinates of the private vertices lie outside the first-coordinate intervals of the primary owners. The residues of a private vertex's first two coordinates modulo $t$ determine $i$ and $b$, so the blocks for distinct routes are disjoint.

Use $x$ as the anchor of the loop at $x\in X$. The cover for row $i$ and exit $b$ omits this loop whenever $x+\mathbf v_{ib}$ lies in a hole. If an owner $x$ of $\mathcal R$ in row $i$ serves as a residual anchor at exit $b$, then $x+\mathbf v_{ib}\notin\mathcal R$. Since $x$ belongs to no other primary owner set and to no private block, $x+\mathbf v_{ib}\in C$.

For each pattern in a row, assign distinct residual anchors to as many shared arcs as possible. Assign distinct external anchors of the same pattern to the remaining arcs. The net demand bounds guarantee enough such anchors. The sets of external anchors in this row are disjoint and lie outside $X$. If a point is used twice as a residual anchor, the two uses occur on disjoint intervals of exits. Its loop is absent throughout both intervals. Each cycle cover uses a single row, so reusing an anchor in another row cannot repeat a tag within that cover. Thus the loops and shared arcs in each cycle cover have distinct anchors, and translation by the same $\mathbf v_{ib}$ gives them distinct tags.

If the route passes through $a$ shared owner vertices, its cycle cover contains $a$ shared arcs and $|X|-a-R$ loops. The shared arcs and loops use $m-1-R$ distinct tags. Assign the remaining $R+1$ tags bijectively to the $R+1$ arcs of its private path. Each such arc occurs only on this route, so subtracting $\mathbf v_{ib}$ gives it a fixed anchor. By \cref{eq:main-exposure,lem:exposed-translations}, this gives precisely $t^2$ edge-disjoint perfect matchings.

\subsection{Rectangle counts and the size of the active support}\label{app:rect-support}

\begin{proof}[Proof of \cref{prop:rectangular-factor}]
We bound the number of primary levels and count the sites available for private holes. We then show that the hub, the primary owners, and the vertices in the private blocks together number at most $m/k$. The choice of $T$ gives $T\ge224$ and $128k^3T^7\le m<128k^3(T+1)^7$. Since $AB=24k^2T^4$ and $(1+1/T)^7\le1+7/T+22/T^2$ for $T\ge100$, the height satisfies
\begin{equation}\label{eq:rect-height}
 \frac{16}{3}kT^3-2\le H\le\frac{16}{3}kT^3+39kT^2 .
\end{equation}
In particular $H-3D\ge5kT^3$ and $Q\ge5kT^2$.

First consider the solution of \cref{eq:rect-calibration}. From $S-A=3AD/(H-3D)$ we get $S\le A+6T/5$. Also $AB/Q\le A+7T/5$, since $(A+7T/5)Q-AB=2kT^3/5+(4k-14/5)T^2+7T/5>0$. These bounds give $D(T-1)\{S+(AB-S)/Q\}\le9kT^5/2$. The denominator defining $z$ is positive, since
\[
 \frac{3D}{H}+\frac{D(T-1)}{QH}
 \le\frac{3}{10kT}+\frac1{50k^2T^2}\le\frac1{3T}.
\]
This denominator is at most $BH$. As $BH\ge30k^2T^5$, this yields $z\ge T-3/(20k)>T-1$. To bound $z$ from above, note that both $B/(B-2T)$ and $H/(H-3D-D(T-1)/Q)$ are at most $1+1/(2T)$. By \cref{eq:rect-height}, $AD/H\le3T/8+9/(40T^2)$. Dropping the subtracted term in the numerator now gives
\[
 z\le\left(1+\frac1{2T}\right)^2
       \left(\frac{11T}{8}+\frac9{40T^2}\right)\le\frac{11T}{8}+2 .
\]
Subtracting $11T/8$ from the middle expression gives a remainder that decreases with $T$ and is less than 2 at $T=2$. Since $S\ge A=4kT^2$ and $z\le11T/8+2$, we have $x=S-z>1$. Since $0<x<S<Q$ and $AB/Q>A$, also $y=(AB-x)/Q-z>A-1-z>2$.

Each horizontal level has room for $L_h=3kT$ rectangles. For a vertical level,
\[
 \frac{H-D}{2D}
 \ge\frac{16kT}{3}+\frac{16kT}{3(T-1)}-\frac12-\frac{2}{T(T-1)}
 >\frac{16kT}{3}+1.
\]
Taking the floor gives $L_v\ge16kT/3$. Using $n_h\le S$, $n_v\le AB/Q$ and $n_f\le z$ in \cref{eq:rect-level-count} gives
\begin{equation}\label{eq:rect-levels}
 U_0\le\left(\frac{4T}{3}+\frac75\right)
      +\left(\frac{3T}{4}+\frac{101}{80}\right)
      +\left(\frac{11T}{8}+3\right)
 \le\frac{83T}{24}+6.
\end{equation}

For private holes, \cref{eq:rect-residue,eq:rect-height} give
\[
 R_*\le\frac{46}{3}kT^5+(24k^2+39k+1)T^4+10kT^3+T^2
      \le\frac{46}{3}kT^5+65k^2T^4.
\]
Rounding reduces the horizontal area by less than $T(H-2D)$ and the vertical area by less than $2(B-T)D$. After accounting for the shorter rectangle, if present, the area lost from the full type is at most $(B-T)(D+1)$. In all, $0\le R-R_*<TH+3BD+B<15kT^4$, the last inequality following from $43/3+39/T+6/T^2<15$. Consequently
\begin{equation}\label{eq:rect-private}
 0<R\le \frac{46}{3}kT^5+80k^2T^4.
\end{equation}
The floor in $P_x$ gives $P_x=4k(T-1)+\lfloor(4k-1)/(T+1)\rfloor$, and likewise $P_y\ge6k(T-1)$. Thus $P_x-U_0\ge k(13T/24-10)>0$, and $H-2D\ge k(16T^3/3-T^2)$. The number of available private points is at least
\[
 k^3\left(\frac{13T}{24}-10\right)(6T-6)
       \left(\frac{16T^3}{3}-T^2\right)
 \ge k^3\left(\frac{52T^5}{3}-341T^4\right).
\]
After factoring out $k^3$, the coefficients of $T^5,T^4,T^3,T^2$ are $52/3,-4087/12,1533/4,-60$, respectively. The lower bound on the number of available sites exceeds the upper bound on $R$ in \cref{eq:rect-private} by at least $k^3T^4(2T-421)>0$. Since $U_0\le P_x=\lfloor(A-1)/t\rfloor$, the primary levels $t,2t,\ldots,U_0t$ all lie in $[T,A-1]$.

Take $U$ to consist of the hub, all primary owners, and all vertices of the private blocks. The owners of rectangles on any one primary level occupy at most $t$ first coordinates, with at most $BH$ points per coordinate in the full layers of $X$. The $t^2$ private blocks each have $R$ vertices. Hence $|U|\le 1+tU_0BH+t^2R$. Every projection vertex outside $U$ has only its loop. For $k=1$ we already have $|U|\le m$. For $k\ge2$, using $m\ge ABH$, $H\ge5kT^3$, and \cref{eq:rect-levels,eq:rect-private}, we obtain
\[
 \frac{k|U|}{m}
 \le \frac{83}{96}+\frac{227}{96T}+\frac{3}{2T^2}
   +\left(1+\frac1T\right)^2\left(\frac{23}{180k}+\frac{2}{3T}\right)
   +\frac1{120k^2T^7}<1 .
\]
Indeed, the first three terms sum to less than $9/10$ for $T\ge100$. The fourth is at most $(101/100)^2(23/360+1/150)<3/40$ and the last is less than $1/40$. This proves the support bound.
\end{proof}

\subsection{A planar construction}\label{app:planar}

\begin{proposition}\label{prop:planar-factor}
Let $k\ge1$ and $2\le T<224^2$ be integers. For every integer $m>128k^3T^{7/2}$, there is a balanced tripartite hypergraph of part size $m$ that is $(T+1)$-rigid and has an active support $U$ satisfying $k|U|\le m$.
\end{proposition}

\begin{proof}
Put $t=T+1$, $\gamma_b=\binom b2$, $D=\gamma_T$, $p=32k$, $w=pt$, and $B=(T-1)^2+1$. The planar translations are $\mathbf v_b=(b,\gamma_b)$ for $0\le b\le T$. The functional $\lambda_b(x,y)=2y-(2b-1)x$ takes its unique minimum among the translations at $\mathbf v_b$, since $\lambda_b(\mathbf v_c-\mathbf v_b)=(c-b)^2$. The projection has one row. Its $t$ routes through $s$ are
\[
 s,S_1,\ldots,S_b,K_b,P_b,\ldots,P_{T-1},s,\qquad 0\le b\le T.
\]
The suffix, prefix, and private chains are as in \cref{app:rectangular}.

Put $R_0=(T^3-T)/2+w-1$. For the box height $h$, we require $h\ge h_0$, where
\begin{equation}\label{eq:planar-height}
 h_0=\max\left\{4D+B,\ 3D+\left\lceil\frac{R_0+D+2B-1}{p-2}\right\rceil\right\}.
\end{equation}
We have $4D+B\le3T^2$ and $p-2\ge30k$. For the second term in the maximum,
\[
 3D+\left\lceil\frac{R_0+D+2B-1}{p-2}\right\rceil
 \le\frac{T^3}{60k}+\frac{19T^2}{12}+\frac{16(T+1)}{15}+1
 \le\frac{T^3}{60k}+3T^2.
\]
Multiplying by $w=32k(T+1)$ gives $wh_0\le(8/15)T^4+145kT^3$. Put $x=\sqrt T$. Since $k\ge1$, the last inequality gives $wh_0/(kT^{7/2})\le 8x/15+145/x$. The function $x\mapsto 8x/15+145/x$ is convex on $[\sqrt2,224]$ and is less than 128 at both endpoints. Consequently $wh_0<128kT^{7/2}\le128k^3T^{7/2}<m$. Write $m-1=wh+\rho$, where $0\le\rho<w$. Then $h\ge h_0$.

Let $X$ consist of $[w]_0\times[h]_0$ and $\rho$ arbitrary points in the next row $[w]_0\times\{h\}$. Use the tag region
\[
 Y=\bigcup_{b=0}^T\bigl([w]_0\times[0,h]+\mathbf v_b\bigr),
 \qquad |Y|=(w+T)(h+1)+(w-1)D .
\]
As $b$ increases, both translation coordinates are nondecreasing. Any point of rectangle $b$ that lies in an earlier rectangle therefore also lies in rectangle $b-1$. Thus the $b$th rectangle adds $(h+1)+(w-1)(\gamma_b-\gamma_{b-1})$ points for $b\ge1$.

\emph{The two sets of primary holes.}
Remove
\[
\begin{aligned}
 Z_1&=\bigl([T,2T-1]\times[D,h-1]\bigr)\cup\bigl([2T,w-1]\times[D,2D-1]\bigr),\\
 Z_2&=[3T,4T-1]\times[3D,3D+B-1].
\end{aligned}
\]
All inverse translates are in the full $w$ by $h$ box. Every inverse translate of $Z_1$ lies in the union of the strips $x<2T$ and $y<2D$. Those of $Z_2$ lie in $[2T,4T-1]\times[2D,3D+B-1]$. The 2 owner sets are therefore disjoint.

For each owner $x$ of $Z_1$, the exits $b$ with $x+\mathbf v_b\in Z_1$ form an interval. Both coordinates of $x+\mathbf v_b$ are nondecreasing in $b$, and at first coordinate $2T$ the largest height allowed in $Z_1$ drops from $h-1$ to $2D-1$. The interval is proper, since a point whose translate at exit 0 lies in $Z_1$ has first coordinate at least $T$ and height at least $D$, so its translate at exit $T$ misses $Z_1$. For either primary set, if the interval is a prefix or a suffix, omit the loop precisely on that interval. If the translate lies in $Z_1$ on an internal interval $[a,b]$, omit the loop throughout $[0,b]$ and use the point as a residual anchor on $[0,a-1]$.

For $1\le j\le T$, consider the anchors required by $Z_1$ alone. After the residual anchors are assigned, the prefix $[0,j-1]$ and the suffix $[j,T]$ each need $\sigma_j$ external anchors, where
\[
 \sigma_j=h-2D+\gamma_j+(w-2T+j-1)(j-1).
\]
To count suffixes starting at $j$, first consider points with first coordinate $T-j$ before translation. Entry through the vertical arm gives $h-2D+\gamma_j$ points. In each of the next $j-1$ columns, exactly $\gamma_j-\gamma_{j-1}=j-1$ points enter across the lower horizontal boundary. The horizontal arm contributes a further $(w-2T)(j-1)$. At each of the exits $j-1$ and $j$, exactly $|Z_1|$ owners have translates in $Z_1$. Hence the number of intervals ending at $j-1$ equals the number starting at $j$. After accounting for the residual anchors on the prefix, the two net demands agree.

For a fixed owner, the $T+1$ translates have distinct first coordinates. Since $Z_2$ has only $T$ columns, the owner's interval in $Z_2$ is proper. If the translate lies in $Z_2$ on an internal interval $[a,b]$, omit the loop throughout $[a,T]$ and use the point as a residual anchor on $[b+1,T]$. The number of points whose tags lie in $Z_2$ at both exits $0$ and $b$ is $I_b\coloneqq |(Z_2-\mathbf v_0)\cap(Z_2-\mathbf v_b)|=(T-b)(B-\gamma_b)$, since $B\ge D$. Thus $Z_2$ contributes $I_{j-1}-I_j=B-\gamma_j+(T-j+1)(j-1)$ to the prefix demand. Since at every exit $b$ exactly $TB$ points $x\in X$ satisfy $x+\mathbf v_b\in Z_2$, the same count gives the net suffix demand. Together the 2 sets require
\[
 E_j=\sigma_j+I_{j-1}-I_j=h-T+2+(w-T)(j-1)
\]
external anchors for each of the two patterns.

\emph{Anchors on the shared chains.}
For the prefix take the right column and top strip, and for the suffix take the left column and bottom strip, as follows.
\[
\begin{array}{@{}l@{\quad}l@{\quad}l@{}}
 \text{Pattern}&\text{Column}&\text{Strip}\\
 \midrule
 {}[0,j-1]&\{w+T-j\}\times[D-\gamma_{j-1},h+D-\gamma_{j-1}]
         &[T,w-1]\times[h+1+D-\gamma_j,h+D-\gamma_{j-1}]\\
 {}[j,T]&\{-j\}\times[-\gamma_j,h-D+\gamma_{T-j}]
      &[0,w-T-1]\times[-\gamma_j,-\gamma_{j-1}-1]
\end{array}
\]
Also use the following lower intervals on the last $T$ columns, where $0\le a<T$.
\[
 \{w-1-a\}\times[-F_a(j),-F_a(j-1)-1],\qquad
 F_a(b)=\gamma_b-\gamma_{\max\{0,b-a\}} .
\]
In column $w-1-a$, the lower boundary of $Y-\mathbf v_b$ is at height $-F_a(b)$. For each column these intervals for different $j$ are disjoint. The bottom strip occupies only earlier columns. All the displayed anchor sets are pairwise disjoint and lie outside $X$.

We check that every allowed translate of these anchors lies in $Y$. For the right-hand column assigned to the prefix $[0,j-1]$, put $q=T-j$. At translated first coordinate $w+q+b$ with $b\le j-1$, the vertical section of $Y$ is $[\gamma_{q+b+1},h+D]$. This section contains every translated anchor in that column because $\gamma_{q+b+1}-\gamma_b$ is nondecreasing in $b$ and $h+D-\gamma_b$ is nonincreasing. For the left-hand column assigned to the suffix $[j,T]$, the section at $b-j$ is $[0,h+\gamma_{b-j}]$. Here $b\ge j$ and $D-\gamma_b\ge \gamma_{T-j}-\gamma_{b-j}$. In each of the last $T$ columns, $-F_a(b)$ is nonincreasing in $b$. Thus the extra lower intervals are valid throughout the suffix, with translated height below $D$. The top strip for the prefix translates into heights $[h+1,h+D]$ at first coordinates at least $T$. The bottom strip for the suffix translates into $[0,w-1]\times[0,D-1]$. An anchor from the right-hand column translates to a first coordinate greater than $w-1$, and one from the left-hand column to a first coordinate less than $T$. Hence all these tags avoid both primary sets and the private holes specified below.

The set of prefix anchors has size $h+1+(w-T)(j-1)=E_j+T-1$. The last $T$ columns add $\sum_{a=0}^{T-1}\min\{j-1,a\}=D-\gamma_{T-j+1}$ suffix anchors. The set of suffix anchors therefore has size
\[
 h+1-j(T-j)+(w-T)(j-1)+D-\gamma_{T-j+1}=E_j+\gamma_{j+1}-1\ge E_j.
\]
After the residual anchors have been assigned, each chain has at least as many external anchors available as there are shared arcs still to assign.

\emph{Private holes and completion.}
The primary holes number $Th+(w-3T)D+TB$. Thus $R=|Y|-m-|Z_1|-|Z_2|=(T^3-T)/2+w-\rho-1=R_0-\rho>0$. Choose $R$ points from the union of the three disjoint sets
\[
\begin{aligned}
 &\{3T\}\times[3D+B,h-1],\\
 &\{4T+1\}\times\bigl(\{3D\}\cup[4D+B,h-1]\bigr),\\
 &\{T-2+at:4\le a<p\}\times[3D,h-1].
\end{aligned}
\]
The union has size $(p-2)(h-3D)-(D+2B-1)\ge R_0$ by \cref{eq:planar-height}. Call the chosen set $Z_K$. The translates $K_b=Z_K-\mathbf v_b$ are pairwise disjoint because their first coordinates have distinct residues modulo $t$. They lie in $X$ and avoid the two strips containing the $Z_1$ owners. For a chosen point $z=(3T,z_2)$, the condition $z-\mathbf v_b\in Z_2-\mathbf v_c$ would force $c\ge b$ and hence $z_2\le3D+B-1$. This contradicts the choice of heights. For a chosen point with first coordinate $4T+1$, the same condition forces $b\ge c+2$, giving $3D+1\le z_2\le4D+B-1$, also excluded. The inverse translates of points in all later columns have first coordinate at least $4T+2$.

Take $C=Y\setminus(Z_1\cup Z_2\cup Z_K)$, of size $m$. Each residual anchor is used at exits where its loop is absent and its translate lies outside the corresponding primary hole set. For each of $Z_1,Z_2,Z_K$, take the union of its translates by $-\mathbf v_b$ over all exits $b$. These three unions are pairwise disjoint, so the residual translate also misses the other holes. The sets of external anchors are disjoint and lie outside $X$, so the assignment by pattern in \cref{app:rectangular} applies.

The positive counts $\sigma_j$ ensure that every shared chain is nonempty. Insert the indicated routes and the loops on $X$. Every directed cycle of length at least 2 passes through $s$ and uses a single private path. In each cycle cover, the shared arcs and loops use $m-1-R$ distinct tags. Assign the remaining $R+1$ tags bijectively to the $R+1$ arcs of the private path, which occur only in that cover. For each $b$, the functional $\lambda_b$ attains its minimum over the translations only at $\mathbf v_b$. Hence \cref{lem:exposed-translations} rules out every other perfect matching.

\emph{Active support.}
Take $U$ to consist of the hub, all primary owners, and all vertices of the private blocks. The union of the two strips for $Z_1$ contains at most $2Th+2D(w-2T)$ owners. The rectangle containing the $Z_2$ owners has $2T(D+B)$ points. The private blocks have $tR$ vertices in all. It follows that
\[
 |U|\le 2Th+2D(w-T)+2TB+tR+1.
\]
Here $2Th<m/(16k)$. Using $R\le R_0$, we bound the sum of the remaining terms by
\[
 \frac{T^4}{2}+\left(32k+\frac32\right)T^3+\left(32k-\frac72\right)T^2
       +\left(32k+\frac52\right)T+32k.
\]
Subtract $T^4/2$ from the preceding polynomial and divide the remainder by $T^3$. The result decreases for $T\ge2$ and equals $60k+3/8\le61k$ at $T=2$. Set $x=\sqrt T$. Since $m>128k^3T^{7/2}$,
\[
 \frac{k|U|}{m}
 <\frac1{16}+\frac{1}{128k^2}\left(\frac{x}{2}+\frac{61k}{x}\right)
 \le\frac1{16}+\frac{1}{128k}\left(\frac{x}{2}+\frac{61}{x}\right)<1.
\]
The convex function $x/2+61/x$ is below $113$ at both ends of $[\sqrt2,224]$, and $1/16+113/(128k)\le121/128$. This proves the proposition.
\end{proof}

\subsection{Completion at every part size}

\begin{proof}[Proof of \cref{thm:growth}]
Recall that $k=r-2$ and $u=(m/(128k^3))^{2/7}$. We need at least $u^k$ perfect matchings. If $u\le1$, a single perfect matching suffices. If $1<u\le2$, then $m\ge2$. Starting with the two perfect matchings of a bipartite $2m$-cycle, the cyclic extension in \cref{sec:third-part} doubles their number for each of the $k$ added parts, giving $2^{k+1}\ge u^k$ matchings.

For $2<u<224^2$, take $T=\lceil u\rceil-1$. Then $2\le T<224^2$, $T<u\le T+1$, and $m>128k^3T^{7/2}$. By \cref{prop:planar-factor,thm:support-product}, the product of $k$ planar factors has $(T+1)^k\ge u^k$ matchings.

For $u\ge224^2$ we have $m\ge128k^3\,224^7$. Take $T=\lfloor\sqrt u\rfloor$. By \cref{prop:rectangular-factor,thm:support-product}, the product of $k$ rectangular factors has $(T+1)^{2k}>u^k$ matchings. All four constructions have exactly $m$ vertices per part.
\end{proof}

\section{The 9-vertex upper bound}\label{app:finite}

To prove the upper bound in \cref{thm:finite-exact}, suppose that a $d$-rigid $3$-graph on 9 vertices exists for some $d\ge14$. The union $H$ of any 14 of its perfect matchings is $14$-rigid, with 42 edges and degree 14 at every vertex.

For a pair $ab$, put $W\coloneqq V(H)\setminus\{a,b\}$ and $X\coloneqq\{x\in W\colon abx\in H\}$. Define the auxiliary graph for this $X$ on all 35 triples of $W$ by joining two triples when they are disjoint and the unique point of $W$ outside their union belongs to $X$. Color each edge by that omitted point. Let $G\coloneqq H\cap\binom W3$. In the auxiliary graph, select the vertices corresponding to the edges of $G$. The subgraph induced by these vertices consists of $|X|$ disjoint edges, one of each color, together with isolated vertices. Indeed, a disjoint pair omitting $x\in X$ extends $abx$ to a perfect matching of $H$, and rigidity gives exactly one such extension. Two of these pairs cannot share a triple, because their perfect matchings would then share an edge.

\subsection{Excluding pair codegrees above 4}

Let $X$ be a 5-element set. We shall bound the number of vertices that can be selected in the following bipartite graph. Its two parts are copies of $\binom X2$. Join a left pair to a right pair when they are disjoint, and color the edge by the unique point of $X$ outside their union.

\begin{claim}\label{clm:five-pair-observation}
Select vertices of this graph so that the induced subgraph has maximum degree at most 1 and its edges have distinct colors. If it has 5 edges, exactly 10 vertices are selected. If it has 4 edges, at most 9 vertices are selected.
\end{claim}
\begin{proof}
Write $(P_i,Q_i)$, $i\in[\ell]$, for the matched left and right pairs, and let $x_i$ be their distinct omitted points. Thus $P_i\cap Q_i=\varnothing$, whereas $P_i\cap Q_j\ne\varnothing$ for $i\ne j$. For $x\in X$, let $\alpha_x$ and $\beta_x$ count its occurrences among the $P_i$ and $Q_i$, respectively. Then
\[
 \ell(\ell-1)\le\sum_x\alpha_x\beta_x.
\]
If $\ell=5$, then $\alpha_x+\beta_x=4$ for every $x$, so the right side is at most $20$. Equality holds throughout. Hence $\alpha_x=\beta_x=2$, and $P_i$ and $Q_j$ meet in exactly one point whenever $i\ne j$. In particular the two families of pairs are disjoint. Each is the edge set of a simple $2$-regular graph on 5 vertices, hence a $5$-cycle, and together they partition $\binom X2$. An additional selected left pair would have to intersect every edge of the right $5$-cycle. This is impossible because a $5$-cycle has no vertex cover of size 2. The same applies on the other side. Thus there are exactly 10 selected vertices.

Suppose next that $\ell=4$, and let $x_*$ be the point not among the 4 omitted points. Now $\alpha_x+\beta_x=3$ for $x\ne x_*$ and $\alpha_{x_*}+\beta_{x_*}=4$. Consequently
\[
 12\le\sum_x\alpha_x\beta_x\le4\cdot2+4=12.
\]
Again $|P_i\cap Q_j|=1$ whenever $i\ne j$. Also $\alpha_{x_*}=\beta_{x_*}=2$, and the other 4 $\alpha_x$ belong to $\{1,2\}$. Since $\sum_x\alpha_x=8$, the left pairs form a graph with degree sequence $(2,2,2,1,1)$. This graph is either a path on 5 vertices or the disjoint union of a triangle and an edge.

We rule out the second case. Each matched right pair $Q_i$ meets exactly three of $P_1,\ldots,P_4$ and is distinct from all four. It would therefore join a triangle vertex to an endpoint of the isolated edge. But then every $Q_i$ would meet that edge, leaving it without a disjoint partner.

We may therefore label $X=[5]$ so that the four matched left pairs are $12,23,34,45$. Each $Q_i$ meets exactly three of them and is distinct from all four. The only possibilities are $13,14,25,35$, so these are the four matched right pairs. An additional selected left pair must meet all four. The only 2-element vertex cover of their path is $15$. Similarly the only possible additional right pair is $24$. These 2 pairs are disjoint, so selecting both creates a fifth edge in the induced graph. At most one can be added, proving the upper bound 9.
\end{proof}

\begin{lemma}\label{lem:four-codegree}
Every pair of vertices in a $14$-rigid $3$-graph on nine vertices has codegree at most four.
\end{lemma}
\begin{proof}
Suppose a pair $a,b$ has codegree $k\ge5$. With $W,X,G$ as above, $|W|=7$, $|X|=k\le7$, and $|G|=42-(14+14-k)=14+k$. In the auxiliary graph for this $X$, the selected vertices induce a subgraph consisting of $k$ disjoint edges, one of each color in $X$, and isolated vertices.

If $k=7$, the full graph on the $35$ triples is $4$-regular. Its selected set $G$ has size $21$ and induces 7 edges. Thus the cut from $G$ to its complement has size $4\cdot21-2\cdot7=70$, but the 14 vertices outside $G$ are incident with at most $4\cdot14=56$ cut edges.

If $k=6$, write $W=X\cup\{y\}$. The auxiliary graph for this $X$ is bipartite. Triples containing $y$ are identified with pairs of $X$, and a triple contained in $X$ is identified with its 3-element complement in $X$. A left pair is adjacent to a right triple exactly when the pair is contained in the triple. The left and right degrees are 4 and 3, respectively. Let $p$ left vertices and $q=20-p$ right vertices be selected. The 6 induced matching edges imply $p\ge6$. Counting edges from the selected left vertices to the unselected right vertices gives $4p-6\le3(20-q)=3p$, so $p=6$. Equality means that each of the 6 unselected right triples has all 3 of its pairs among the 6 selected left pairs. The graph on $X$ formed by the six selected pairs would therefore contain at least six triangles. But each triangle uses 3 pairs of adjacent edges, and each pair of edges belongs to at most one triangle. Hence a graph with 6 edges has at most $\binom62/3=5$ triangles, a contradiction.

It remains to consider $k=5$. Write $W=X\cup\{u,v\}$. The auxiliary graph for this $X$ has two components. The first has triples $uvx$ on one side and triples contained in $X$ on the other. Identifying each triple in $X$ with its complementary pair gives the vertex--edge incidence graph of $K_5$. The second has triples $uP$ and $vQ$, with $P,Q\in\binom X2$, and joins them when $P\cap Q=\varnothing$. It is the graph of \cref{clm:five-pair-observation}.

Let $s$ of the 5 induced matching edges lie in the incidence component. Let $A\subseteq X$ be the set of vertices of $K_5$ selected on the left of its incidence graph. Each selected vertex on the other side represents an edge of $K_5$ with at most one endpoint in $A$. Exactly $s$ of these edges have an endpoint in $A$, and the others lie in $X\setminus A$. Thus the number $g_1$ of selected vertices in this component satisfies
\[
 g_1\le |A|+\binom{5-|A|}{2}+s.
\]
If a selected incidence edge joins $x\in A$ to the edge $\{x,y\}$ of $K_5$, its color is $y\in X\setminus A$. Since the $s$ colors are distinct and the induced edges form a matching, $s\le\min\{|A|,5-|A|\}\le2$.

The other component is a $3$-regular graph on 20 vertices. If it contains $g_2$ selected vertices and $\ell=5-s$ induced edges, counting the edges from the $g_2$ selected vertices to the $20-g_2$ unselected vertices gives
\[
 3g_2-2\ell\le3(20-g_2),
 \quad\text{and hence}\quad g_2\le10+\left\lfloor\frac\ell3\right\rfloor.
\]
If $s=2$, then $|A|\in\{2,3\}$, so $g_1\le7$ and $g_2\le11$, contradicting $g_1+g_2=|G|=19$. If $s=1$, then $g_1\le8$, while \cref{clm:five-pair-observation} gives $g_2\le9$, again a contradiction.

Thus $s=0$. The same claim gives $g_2=10$, and hence $g_1=9$. The bound for $g_1$ forces $A=\varnothing$. Consequently $G$ consists of 9 of the 10 triples contained in $X$, together with 10 triples of the form $uP$ or $vQ$. These ten triples are the endpoints of the five induced edges, so they are all used in the 5 perfect matchings containing $abx$, $x\in X$.

There are 9 remaining perfect matchings of $H$. Each contains one of the 9 triples of $X$, one edge containing $a$ but not $b$, and one edge containing $b$ but not $a$. There is at most one edge of the form $axy$ with $x,y\in X$, because any such edge must be paired with the unique possible edge $buv$ in its perfect matching. Similarly there is at most one edge $bxy$ with $x,y\in X$. There is also at most one edge $auv$ and at most one edge $buv$. Hence at least 7 of the 9 edges containing $a$ but not $b$ have the form $aux$ or $avx$ for some $x\in X$.

After interchanging $u$ and $v$ if necessary, we may assume that at least 4 edges have the form $aux$. They belong to distinct perfect matchings. Each of these matchings pairs its edge $aux$ with an edge $bvy$, and the latter edges are distinct by rigidity.

Fix one such $aux$. There are at least four choices of $y$ with $bvy\in H$. At most one equals $x$, and at most one other choice makes $X\setminus\{x,y\}$ the unique triple of $X$ missing from $G$. Each remaining $y$ gives a perfect matching $\{aux,bvy,X\setminus\{x,y\}\}$. At least 2 of them therefore contain $aux$, contradicting rigidity.
\end{proof}

\subsection{Triples disjoint from a pair of codegree four}

\begin{lemma}\label{lem:four-neighborhood}
Suppose that $H$ is a $14$-rigid $3$-graph on 9 vertices and that every pair has codegree at most 4. There is a pair $ab$ of codegree 4 for which one can label the other 7 vertices $0,1,\ldots,6$ so that $\{x\colon abx\in H\}=\{0,1,2,3\}$. With this labeling, the edges disjoint from $\{a,b\}$ are
\[
\begin{aligned}
 \{&012,013,014,015,023,024,026,034,035,046,056,\\
     &123,125,136,146,236,245,345\}.
\end{aligned}
\]
Here, for example, $012$ denotes $\{0,1,2\}$.
\end{lemma}
\begin{proof}
Since the sum of the 36 pair codegrees is $3|H|=126>3\cdot36$, some pair $ab$ has codegree 4. With $W,X,G$ as defined above for this pair, we have $|X|=4$ and $|G|=18$. Put $Y\coloneqq W\setminus X$. In the auxiliary graph for $X$, the selected vertices induce a subgraph with four disjoint edges, one of each color in $X$.

For $y\in Y$, there is no edge $aby$. At most 4 edges contain $a,y$, and at most 4 contain $b,y$, because pair codegrees are at most 4. Hence
\begin{equation}
 \deg_G(y)\ge14-4-4=6\qquad(y\in Y).
 \label{eq:four-G-minimum-degrees}
\end{equation}

Write $Y=\{y_1,y_2,y_3\}$. The auxiliary graph on the 35 triples of $W$ has four components. One is a star $K_{1,4}$, with center $Y$ and leaves $\binom X3$. Index the three remaining components by $i\in[3]$. Component $i$ has left vertices $(Y\setminus\{y_i\})\cup\{x\}$, $x\in X$, and right vertices $\{y_i\}\cup P$, $P\in\binom X2$. Identifying each right vertex $\{y_i\}\cup P$ with the pair $X\setminus P$ gives the vertex--edge incidence graph of $K_4$. Its incidence edge joining a vertex $x$ to a pair $\{x,x'\}$ has color $x'$.

Let $a_i,b_i,s_i$ count the selected left vertices, selected right vertices, and induced matching edges in the $i$th incidence component. In $K_4$, each selected edge has at most one selected endpoint. Exactly $s_i$ selected edges have such an endpoint. The other selected edges lie among the $4-a_i$ unselected vertices. Consequently
\begin{equation}
 b_i\le\binom{4-a_i}{2}+s_i,
 \qquad s_i\le\min\{a_i,4-a_i\}\le2.
 \label{eq:four-component-bound}
\end{equation}
The $s_i$ induced edges have distinct left endpoints among the $a_i$ selected left vertices. Their colors are distinct points among the other $4-a_i$ vertices of $X$. For $s_i=0,1,2$ the maximum numbers $a_i+b_i$ of selected vertices are $6,5,5$, respectively. Attaining these maxima forces $(a_i,b_i)=(0,6),(1,4),(2,3)$, respectively. If $s_i=2$, then $a_i=2$ and $b_i\in\{2,3\}$.

Let $\varepsilon$ indicate whether the star center $Y$ is selected, and let $\ell$ count its selected leaves. Directly from the component descriptions,
\begin{equation}
 \deg_G(y_i)=\varepsilon+b_i+a_j+a_k,
 \qquad \{i,j,k\}=\{1,2,3\}.
 \label{eq:four-Y-degrees}
\end{equation}
By \cref{eq:four-G-minimum-degrees}, every distribution must satisfy $\varepsilon+b_i+a_j+a_k\ge6$ whenever $\{i,j,k\}=\{1,2,3\}$.

If the star supplies one of the 4 induced matching edges, then $\varepsilon=\ell=1$. Up to reordering the three incidence components, $(s_1,s_2,s_3)$ is $(2,1,0)$ or $(1,1,1)$. The latter permits at most $2+5+5+5=17$ selected vertices. In the former, 18 selected vertices force every component to attain its maximum, giving $(a_i,b_i)=(2,3),(1,4),(0,6)$. At $y_1$, \cref{eq:four-Y-degrees} gives $\deg_G(y_1)=1+3+1+0=5$, contrary to \cref{eq:four-G-minimum-degrees}.

Thus the star contains no induced matching edge. If its center were selected, it would be the only selected vertex of the star. The three incidence components together contain at most $5+5+6=16$ selected vertices, giving fewer than 18 in all. Hence $\varepsilon=0$ and $0\le\ell\le4$. The 4 matching edges are distributed among the incidence components as $(2,1,1)$ or $(2,2,0)$.

First suppose $(s_1,s_2,s_3)=(2,1,1)$. Each incidence component contains at most 5 selected vertices, so $\ell\in\{3,4\}$. If $\ell=3$, all three contain 5 selected vertices, and \cref{eq:four-Y-degrees} gives $\deg_G(y_1)=3+1+1=5$. If $\ell=4$ and the first component contains 4 selected vertices, the other two contain 5 each, giving $\deg_G(y_1)=2+1+1=4$. Both cases contradict \cref{eq:four-G-minimum-degrees}. Otherwise, after interchanging components 2 and 3 if necessary, the numbers of selected vertices are $5,5,4$. The corresponding $(a_i,b_i)$ are $(2,3),(1,4),(a,b)$ with $a+b=4$. Applying \cref{eq:four-Y-degrees} at $y_1$ and $y_3$ and using the lower bound in \cref{eq:four-G-minimum-degrees} gives $3+1+a\ge6$ and $b+2+1\ge6$. Thus $a\ge2$ and $b\ge3$, contradicting $a+b=4$.

The remaining distribution is $(2,2,0)$. Number the components so that $s_3=0$. The first two have $a_1=a_2=2$ and at most 5 selected vertices each, so the third contains at least 4 selected vertices. By \cref{eq:four-component-bound}, the possible pairs $(a_3,b_3)$ are $(0,b)$ with $4\le b\le6$, $(1,3)$, and $(4,0)$. The case $a_3=0$ gives $\deg_G(y_1)\le3+2+0=5$, and $(4,0)$ gives $\deg_G(y_3)=0+2+2=4$. Therefore $(a_3,b_3)=(1,3)$. Counting the 18 selected vertices now forces
\begin{equation}
 \ell=4,\qquad
 (a_1,b_1)=(a_2,b_2)=(2,3),\qquad
 (a_3,b_3)=(1,3).
 \label{eq:four-forced-profile}
\end{equation}

Let $P,Q\subseteq X$ consist of the labels of the selected left vertices in the first and second components, respectively. By \cref{eq:four-forced-profile}, each has size 2. The colors of the two matching edges in the first component are the two points of $X\setminus P$, and those in the second are the two points of $X\setminus Q$. Since all 4 colors are distinct, $P$ and $Q$ partition $X$. The 3 selected triples of the form $y_1xx'$ in the first component are $\{y_1\}\cup P$ and the 2 triples obtained by adjoining $y_1$ to the edges of a perfect matching between $P$ and $Q$. The second component likewise gives $\{y_2\}\cup Q$ and 2 triples from a perfect matching between $P$ and $Q$.

Write $x_0$ for the label of the selected left vertex in the third component. This component contributes $y_1y_2x_0$ and the three triples $y_3x_0x$ with $x\in X\setminus\{x_0\}$. The matchings between $P$ and $Q$ chosen in the first and second incidence components must be distinct. Indeed, if they coincide, let $x$ be the partner of $x_0$ in this matching. The pair $x_0x$ then lies in 2 triples of $\binom X3$, in $y_3x_0x$, and in both $y_1x_0x$ and $y_2x_0x$, giving codegree at least 5.

Interchanging the first two incidence components if necessary, assume $x_0\in P$. Relabel so that $P=\{0,1\}$, $Q=\{2,3\}$, $x_0=0$, $y_1=5$, $y_2=6$, and $y_3=4$. Choose the labels $2,3$ in $Q$ so that the first matching is $\{03,12\}$. The second is then $\{02,13\}$. Applying this labeling to the triples described above gives the 18 edges in the statement.
\end{proof}

\subsection{The seven-vertex configuration has no rigid extension}

\begin{lemma}\label{lem:four-normal-form-extension}
There is no $14$-rigid $3$-graph $H$ on $\{a,b,0,1,\ldots,6\}$ with maximum pair codegree at most 4 for which $\{x\colon abx\in H\}=\{0,1,2,3\}$ and $G\coloneqq H[\{0,1,\ldots,6\}]$ has the 18 edges listed in \cref{lem:four-neighborhood}.
\end{lemma}
\begin{proof}
For a vertex $v$ and a pair $S$, write $vS=\{v\}\cup S$. Let $\mathcal A$ consist of the pairs $S\subseteq\{0,\ldots,6\}$ with $aS\in H$, and define $\mathcal B$ likewise for $b$. Both families have size 10, since $H$ is $14$-regular and precisely 4 edges contain both $a$ and $b$.

The perfect matchings through $ab1$, $ab2$, $ab3$, and $ab0$ use, in that order, the following pairs of edges of $G$
\[
 \{026,345\},\quad \{035,146\},\quad
 \{046,125\},\quad \{136,245\}.
\]
These 8 edges of $G$ cannot occur in any further perfect matching. Every remaining perfect matching has the form $\{aS,bT,W\setminus(S\cup T)\}$, where $S\in\mathcal A$, $T\in\mathcal B$, $S\cap T=\varnothing$, and $W\setminus(S\cup T)\in G$.

The pairs $01,02,03,04$ already have codegree 4 in $G$, while $05$ and $06$ have codegree 3. Also $\deg_G(0)=11$, and $ab0\in H$. Consequently $05$ and $06$ each belong to exactly one of $\mathcal A,\mathcal B$, and neither family contains any other pair through $0$. Interchanging $a,b$ if necessary, assume $06\in\mathcal A$.

For $S=06$, the candidate pairs for $T$ are $12,13,34,45$. The first three give the already used triples $345,245,125$, respectively. Thus $45\in\mathcal B$, and the matching containing $a06$ is $\{a06,b45,123\}$. Similarly, the possible partners of $05$ are $14,23,24,46$, with complementary triples $236,146,136,123$, respectively. The last three triples have already been used, so $05$ must be paired with $14$ in the opposite family.

If $05\in\mathcal A$, then $\mathcal B$ contains no pair through $0$. The unique perfect matching containing $a06$ uses $b45$. If $\mathcal B$ contained a pair $T\in\{12,13,34\}$, then $a06$, $bT$, and $W\setminus(\{0,6\}\cup T)$ would form a second perfect matching through $a06$. The same argument for $a05$, whose matching uses $b14$, excludes $23,24,46$ from $\mathcal B$. These are 6 distinct pairs among the 15 pairs on $\{1,\ldots,6\}$, so $|\mathcal B|\le15-6=9$, contradicting $|\mathcal B|=10$.

If $05\in\mathcal B$, then $06$ is the only pair of $\mathcal A$ through $0$. The matching containing $b05$ uses $a14$, so $23,24,46\notin\mathcal A$. The unique perfect matching through $b45$ uses $a06$, so $13,16,26,36\notin\mathcal A$, because their complementary triples with $45$ are $026,023,013,012$, all in $G$. These are 7 distinct pairs on $\{1,\ldots,6\}$, whence $|\mathcal A|\le1+(15-7)=9$, contradicting $|\mathcal A|=10$.
\end{proof}

\begin{proof}[Proof of \cref{thm:finite-exact} (upper bound for $f_3(9)$)]
By \cref{lem:four-codegree,lem:four-neighborhood}, any $14$-rigid $3$-graph on 9 vertices would contain the configuration excluded by \cref{lem:four-normal-form-extension}. Hence $f_3(9)\le13$. The constructions and the tripartite upper bound in \cref{sec:finite} complete the proof.
\end{proof}

\bibliographystyle{abbrv}
\bibliography{references}

\clearpage
\pdfbookmark[0]{Supplemental Material}{supplemental-material}
\section*{Supplemental Material}
\fi
\setcounter{section}{0}
\renewcommand{\thesection}{S\arabic{section}}
\renewcommand{\theHsection}{\thesection}
\crefalias{appendix}{section}
\crefalias{subappendix}{subsection}
\setcounter{table}{0}
\renewcommand{\thetable}{S\arabic{table}}
\renewcommand{\theHtable}{\thetable}

\ifstandalonesupplement
  A hypergraph is $d$-rigid if it is the union of $d$ pairwise edge-disjoint perfect matchings and has no others. Each perfect matching in an $r$-graph on $n=rm$ vertices has $m$ edges. Write $f_r(n)$ for the largest possible $d$, and $g_r(m)$ for the maximum in a balanced $r$-partite $r$-graph with $m$ vertices per part.\par
\fi

We give explicit examples of small networks and prove that grafting and certain cyclic lifts preserve rigidity. For triple systems, let $\ell_3(m)$ be the largest $d$ for a linear $d$-rigid tripartite $3$-graph with $m$ vertices in each part. In a linear hypergraph any two distinct edges share at most one vertex. For the dimensions listed in \cref{finite:sec:thresholds}, we give thresholds $m_0$ such that the corresponding dimension is attainable at every matching size $m\ge m_0$.

\subsection*{Contents}
\startcontents[supplement]
\printcontents[supplement]{supp}{1}[1]{}

\section{Elementary constructions at every part size}\label{finite:sec:elementary}
For $r=3$, the three classes of hypergraphs are nested, so $\ell_3(m)\le g_3(m)\le f_3(3m)$. A $t$-rigid $r$-graph on $rm$ vertices is $t$-regular and has $tm$ edges.
The union of any nonempty subfamily of the prescribed matchings has precisely those perfect matchings.

A bipartite $2m$-cycle has two perfect matchings. The cyclic extension in \cref{sec:third-part} therefore gives $g_r(m)\ge2^{r-1}$ for $m\ge2$. At $m=2$ equality holds, since the $2^r$ transversal edges form complementary pairs.

\begin{theorem}\label{finite:thm:six}
For every $m\ge3$, there is a $6$-rigid tripartite $3$-graph with $m$ vertices in each part, $6m$ edges, and maximum pair codegree at most two. Consequently
\[
 g_3(m)\ge6\quad(m\ge3),\qquad
 g_r(m)\ge3\cdot2^{r-2}\quad(r\ge3, m\ge3).
\]
\end{theorem}
\begin{proof}
Index each of the three parts by $\mathbb Z_m$. If $m$ is odd, put
$\tau(i)=i+1$. If $m$ is even, let $\tau$ be the cycle $(0,1,\ldots,m-1)$
with the entries $1$ and $2$ interchanged.
The following formulas, with $c=0,1$, define six perfect matchings
\begin{align*}
 M_{-,c}&=\{(i,i-1,\tau^c(i)):i\in\mathbb Z_m\},\\
 M_{0,c}&=\{(i,i,\tau^c(i-1)):i\in\mathbb Z_m\},\\
 M_{+,c}&=\{(i,i+1,\tau^c(i+1)):i\in\mathbb Z_m\}.
\end{align*}
For each of the offsets $-1,0,1$ and each $c$, the second and third coordinates, as functions of $i$, are permutations of $\mathbb Z_m$. In the row with first coordinate $i$, the three possible second coordinates are distinct since $m\ge3$. For each of them the two possible third coordinates differ because $\tau$ has no fixed point. Thus the six perfect matchings are pairwise edge-disjoint.

Let $P$ be any perfect matching. Its projection onto the first two parts pairs $i$ with $\sigma(i)$, where $\sigma$ is a permutation satisfying $\sigma(i)\in\{i-1,i,i+1\}$.
Unless $\sigma$ is one of the two cyclic shifts $i\mapsto i\pm1$, each nontrivial cycle of $\sigma$ transposes two neighbors in the cyclic order on $\mathbb Z_m$. Indeed, a cycle of length at least three in the underlying cycle graph must visit every vertex.

At $c=0$, the forward cycle, the backward cycle, and the identity give third coordinates $i+1$, $i$, and $i-1$ in row $i$, respectively. In each case these coordinates run through $\mathbb Z_m$. If the row with unshifted coordinate $z$ uses $\tau(z)$, the row with unshifted coordinate $\tau(z)$ must use $\tau^2(z)$. Since $\tau$ is an $m$-cycle, all rows use the same $c$.

It remains to exclude an adjacent transposition $i\leftrightarrow i+1$. The two corresponding edges have the same third coordinate $i+1$ at $c=0$, so they must use both $i+1$ and $\tau(i+1)$. If $\sigma$ fixed $i+2$, that row would need one of the same two third-part vertices. Since $\sigma$ already transposes $i$ and $i+1$, it must transpose $i+2$ and $i+3$. Continuing around the cycle forces alternating adjacent transpositions. This is impossible for odd $m$. For even $m$, the third coordinates shared by the candidate edges of each transposed pair at $c=0$ form either the set $E$ of even residues or the set $O$ of odd residues. The pairs $\{z,\tau(z)\}$ for these coordinates cover the third part exactly once only if $\tau(E)=O$ or $\tau(O)=E$. Neither equality holds, since $\tau(0)=2$ and $\tau$ is a permutation. Thus $P$ is one of the six prescribed perfect matchings.

A pair in the first and second parts determines the row and the offset, leaving only the two choices for $c$. For a fixed first coordinate, the three possible third coordinates at $c=0$ are distinct. To obtain a given third coordinate $z$, the value at $c=0$ must be $z$ or $\tau^{-1}(z)$, so a first--third pair lies in at most two edges. The three values at $c=0$ are also distinct for a fixed second coordinate, giving the same bound for second--third pairs. This construction gives $g_3(m)\ge6$. Iterating the cyclic extension gives $g_r(m)\ge6\cdot2^{r-3}$ for $r\ge3$.
\end{proof}

\subsection{Eight perfect matchings over \texorpdfstring{$\mathbb Z_m$}{Z/mZ}}\label{finite:subsec:modular-eight}
\begin{proposition}\label{finite:prop:eight-coprime}
If $m\ge5$ and $\gcd(m,6)=1$, then $g_3(m)\ge8$.
\end{proposition}
\begin{proof}
Work modulo $m$. For $h\in\{0,1\}$ and $s\in\{0,1,2,3\}$, put
\[
 M_{h,s}=\{(a,a+s,h-2a-s):a\in\mathbb Z_m\}.
\]
Each $M_{h,s}$ is a perfect matching, and the matchings are pairwise edge-disjoint. Sum the three vertex coordinates over all edges of an arbitrary perfect matching. Since $m$ is odd, the sum in each part is zero modulo $m$. Each edge contributes $h$, so the number $u$ of edges with $h=1$ satisfies $u\equiv0\pmod m$. As $0\le u\le m$, all edges have the same $h$.

Fix a perfect matching with common edge sum $h$ and let $\delta_a\in\{0,1,2,3\}$ be its shift in row $a$. Both maps $a\mapsto a+\delta_a$ and $a\mapsto 2a+\delta_a$ are permutations of $\mathbb Z_m$. Summing the values of the first map gives $\sum_a\delta_a\equiv0\pmod m$. Since $\gcd(m,6)=1$, the formula for $\sum_{a=0}^{m-1}a^2$ gives $\sum_a a^2\equiv0\pmod m$. Summing the squares of the values of each map gives
\[
 2\sum_a a\delta_a+\sum_a\delta_a^2\equiv0\pmod m,\qquad
 4\sum_a a\delta_a+\sum_a\delta_a^2\equiv0\pmod m.
\]
Subtracting the first congruence from the second gives $2\sum_a a\delta_a\equiv0\pmod m$, so the first congruence gives $\sum_a\delta_a^2\equiv0\pmod m$. Let $n_s$ count the rows with shift $s$. Using $\sum_s n_s=m$, $\sum_s s n_s\equiv0$ and $\sum_s s^2 n_s\equiv0\pmod m$, we obtain
\[
 (n_0,n_1,n_2)\equiv n_3(-1,3,-3)\pmod m.
\]
If $n_3\not\equiv0\pmod m$, all four counts lie between $1$ and $m-1$, since $3$ is invertible modulo $m$. Since $n_0\equiv-n_3$ and $n_1\equiv-n_2$ modulo $m$, each pair sums to $m$, contradicting $\sum_s n_s=m$. Thus $n_3\equiv0\pmod m$, and every $n_s$ is a multiple of $m$. Exactly one equals $m$, so the shift is constant.
\end{proof}

\section{Periodic constructions}\label{finite:sec:periodic}
Fix a positive period $q$. For each edge $e$ of a base hypergraph $F$, choose a proper subset $O_e\subsetneq e$. In layer $j\in\mathbb Z_q$, keep the vertices of $e\setminus O_e$ in layer $j$ and shift those of $O_e$ to layer $j+1$. Taking all $q$ lifts of the edges in a base perfect matching gives a perfect matching of the lift.

Identify each layer with $V(F)$. Let $D_{S,T}$ count the choices of edges starting in layer $j$ that cover exactly $V(F)\setminus S$ there and exactly $T$ in layer $j+1$, each vertex once. The edges starting in layer $j+1$ must then leave precisely $T$ uncovered there. Perfect matchings correspond to closed walks of length $q$ starting in layer zero, counted with the multiplicities in $D$. Thus there are $\operatorname{tr}(D^q)$ perfect matchings.

The perfect matchings of the triples (row, column, symbol) of a Latin square are its transversals. McKay, McLeod, and Wanless~\cite{McKayMcLeodWanless2006} enumerated the transversals of all Latin squares through order 9. For a linear tripartite example, take the Latin square of order 8 in entry 16 of Wanless's file \texttt{LS8alphan}~\cite{WanlessTransversalData}. Its eight transversals partition its cells~\cite{EganWanless2012}. Use the shifts in the \ghzdatafile{retained_v11/prior_v10/prior_v9/prior_v8/verification/prior_periodic_certificate.json}{periodic construction data}. The transfer digraph has eight loops of weight one. Its only other directed cycle has length two and shares no vertex with a loop. For odd $q$, the only closed walks of length $q$ therefore stay at a loop. The lifted transversals give all eight perfect matchings. The lift is linear because each pair of base vertices belongs to at most one edge and the lifts of one edge are disjoint. Hence
\begin{equation}\label{finite:eq:latin-odd}
 \ell_3(8q)\ge8\qquad(q\text{ odd}).
\end{equation}

In the nine-vertex example of \texttt{discrete\_things}~\cite{DiscreteThings2024}, omit the two matchings containing $\{1,5,6\}$ and $\{1,6,8\}$, respectively. For the shifts in the \ghzdatafile{retained_v11/prior_v10/prior_v9/prior_v8/prior_results/eleven_certificate.json}{certificate}, the transfer digraph has 11 loops of weight one. The listed vertex heights increase strictly along every other arc, so $\operatorname{tr}(D^q)=11$ and $f_3(9q)\ge11$ for every $q\ge1$.

\section{Grafting and maximum \texorpdfstring{$j$}{j}-degrees}\label{finite:sec:grafting}
For $1\le j\le r$, define the \emph{maximum $j$-degree}
\[
 \Delta_j(H)=\max_{X\in\binom{V(H)}j}d_H(X),\qquad
 d_H(X)=|\{e\in E(H):X\subseteq e\}|.
\]
Thus $\Delta_1$ is the maximum vertex degree and $\Delta_2$ is the maximum pair codegree. A $3$-graph is linear exactly when $\Delta_2\le1$.

\begin{theorem}[Grafting]\label{finite:thm:graft}
Let $H$ be a $t$-rigid $r$-graph on $rm$ vertices, and let $S$ be a $t$-rigid $r$-partite $r$-graph with $q$ vertices in each part.
There is a $t$-rigid $r$-graph $G$ on $r(m+q-1)$ vertices satisfying
\begin{equation}\label{finite:eq:graft-degree}
 \Delta_j(G)\le\max\{\Delta_j(H),\Delta_j(S)\}
 \qquad(1\le j\le r).
\end{equation}
If $H$ is balanced $r$-partite, $G$ can also be chosen balanced
$r$-partite. Consequently
\begin{align}
 f_r(r(m+q-1))&\ge\min\{f_r(rm),g_r(q)\},\label{finite:eq:graft-f}\\
 g_r(m+q-1)&\ge\min\{g_r(m),g_r(q)\}.
 \label{finite:eq:graft-g}
\end{align}
\end{theorem}
\begin{proof}
Write $M_1,\ldots,M_t$ for the prescribed perfect matchings of $H$, and $N_1,\ldots,N_t$ for those of $S$. Let $A_1,\ldots,A_r$ be the parts of $S$. Fix $a_*\in A_1$ and let $e_i$ be the unique edge of $N_i$ through $a_*$. Choose any $r$ distinct vertices $v_1,\ldots,v_r$ of $H$ and replace each $v_j$ by the part $A_j$ of $S$. For each $i$ and each edge of $M_i$, replace every $v_j$ it contains by the unique vertex in $e_i\cap A_j$. Call the images of the edges of $H$ the \emph{inherited edges}. Add the edges of $N_i\setminus\{e_i\}$ as \emph{internal edges} of color $i$. Together with the inherited edges from $M_i$, they cover every vertex exactly once and thus form a perfect matching.

Contracting each $A_j$ to $v_j$ recovers the original edge of $H$ from each inherited edge, so distinct edges of $H$ have distinct images. Every internal edge meets $A_1\setminus\{a_*\}$, and no inherited edge does. The internal edges are distinct because the $N_i$ are edge-disjoint in the simple hypergraph $S$. Thus $G$ is simple and its prescribed perfect matchings are pairwise edge-disjoint.

Let $P$ be any perfect matching of $G$. Only internal edges can cover the $q-1$ vertices of $A_1\setminus\{a_*\}$, and each internal edge covers one of them. The $q-1$ internal edges of $P$ therefore leave exactly one vertex in each $A_j$ for the inherited edges. Contracting those inherited edges gives a perfect matching of $H$, necessarily $M_i$ for some $i$. Together, the inherited edges of $P$ cover precisely the vertices of $e_i$ in $V(S)$. Adjoining $e_i$ to the internal edges of $P$ gives a perfect matching of $S$ containing $e_i$, which must be $N_i$. Hence $P$ is the prescribed perfect matching of color $i$.

If $H$ is balanced $r$-partite, choose $v_j$ in its $j$th part to preserve the given partition. The new $j$th part consists of the old part with $v_j$ replaced by $A_j$ and has size $m+q-1$.

To prove \cref{finite:eq:graft-degree}, fix a $j$-element set $X\subseteq V(G)$ with $j\ge1$. If $X$ contains a retained vertex of $H$, only inherited edges can contain $X$. If $X$ also contains two vertices of the same $A_i$, no such edge exists. Otherwise contraction sends $X$ to a $j$-element subset of $V(H)$ and sends distinct inherited edges containing $X$ to distinct edges of $H$ containing its image. Thus $d_G(X)\le\Delta_j(H)$.

Suppose instead that $X\subseteq V(S)$. An inherited edge of color $i$ can contain $X$ only if $X\subseteq e_i$, and at most one inherited edge of that color can contain the nonempty set $X$. The internal edges containing $X$ are exactly the edges of $S$ containing $X$, except for the deleted edges $e_i$. Each such deleted edge can therefore be replaced by at most one inherited edge containing $X$. Summing over $i$ gives $d_G(X)\le d_S(X)\le\Delta_j(S)$. To obtain \cref{finite:eq:graft-f,finite:eq:graft-g}, let $d_1,d_2$ be the two matching counts and retain $\min\{d_1,d_2\}$ prescribed matchings in each hypergraph before grafting.
\end{proof}

Grafting balanced $t$-rigid hypergraphs of part sizes $q_1,\ldots,q_s$ gives part size $1+\sum_{i=1}^s(q_i-1)$. Any common upper bound on their pair codegrees holds after grafting. In particular, grafting preserves linearity.

\section{Cyclic lifts of finite affine constructions}\label{finite:sec:modular-lifts}
\subsection{The finite seed hypothesis}\label{finite:subsec:affine-seeds}
Let $m$ be odd, and assume that $\lambda$ and $\lambda+1$ are units
modulo $m$. For $S\subseteq\mathbb Z_m$, consider the edge sets
\begin{equation}\label{finite:eq:single-sum}
 P_s=\{(a,\lambda a+s,-(\lambda+1)a-s):a\in\mathbb Z_m\},
 \qquad s\in S.
\end{equation}
They are pairwise edge-disjoint perfect matchings and their union is
linear because any two coordinates determine the third from their zero sum.
The seed hypothesis used below is that the $P_s$, $s\in S$, are the
only perfect matchings of their union.

The \ghzdatafile{certificates.md}{affine seed parameters} specify $m,\lambda,S$ for each seed. For $m=5,7$, restricting \cref{finite:prop:eight-coprime} to $h=0$ proves the seed hypothesis. At $m=3$, the only permutations $b$ of $\mathbb Z_3$ for which $a\mapsto a+b(a)$ is also a permutation are the translations. Deciding whether a tripartite triple system has a perfect matching is Karp's three-dimensional matching problem~\cite{Karp1972}. For the remaining seeds we enumerate all such matchings by exact cover. Row $(a,s)$ covers columns $A_a$, $B_{\lambda a+s}$, and $C_{-(\lambda+1)a-s}$. Branching on one uncovered column and trying each compatible row containing it enumerates every perfect matching once~\cite{KnuthDancingLinks}. For $m=13$ and $\lambda=2$, the \ghzdatafile{retained_v11/prior_v10/prior_v9/verification/finite_slope_checks.json}{enumeration} finds precisely the six perfect matchings $P_s$ with $s\in\{0,1,2,4,5,8\}$.

\begin{lemma}[Doubling a constant-sum seed]\label{finite:lem:double-sum}
Let $m\ge3$ be odd and let $\delta$ be a unit modulo $m$. Suppose the
matchings $P_s$ in \cref{finite:eq:single-sum} are the only perfect
matchings of their union. Then the sets
\[
 N_{h,s}=\{(a,\lambda a+s,\delta h-(\lambda+1)a-s):a\in\mathbb Z_m\},
 \quad h\in\{0,1\},\ s\in S,
\]
are perfect matchings. Their union is $2|S|$-rigid and has maximum
pair codegree at most two.
\end{lemma}
\begin{proof}
For fixed $h$ and $s$, the second and third coordinates are bijective functions of $a$ because $\lambda$ and $\lambda+1$ are units. An edge determines $s$ from its first two coordinates and $h$ from its coordinate sum, so the matchings are edge-disjoint. Sum the three vertex coordinates over all edges of any perfect matching. The sum in each part is zero modulo $m$ since $m$ is odd. Thus $\delta u\equiv0\pmod m$, where $u$ counts the edges with $h=1$. Since $\delta$ is a unit and $0\le u\le m$, all edges have the same $h$. Subtracting $\delta h$ from the third coordinate gives a perfect matching of the original rigid hypergraph. Each of the two hypergraphs obtained by fixing $h$ is linear, so every pair lies in at most two edges of their union.
\end{proof}

For the seed at $m=13$, take $\delta=1$. The two edge sums in \cref{finite:lem:double-sum} give six matchings each, so the \ghzdatafile{examples/twelve_on_39.json}{39-path network} has exactly twelve perfect matchings.

\subsection{Identifying the shift from its bits}
The next lemma gives conditions under which the total multiplicity and one congruence force every marked coordinate to occur once.

\begin{lemma}[Recovery of marked multiplicities]\label{finite:lem:marked-counts}
Let $D\subseteq\{0,\ldots,m-1\}$ be empty or contain $0$. Assume that
$\sum_{a\in D}a<m$ and that each positive element of $D$ exceeds the
sum of all smaller elements of $D$. Let $n_0,\ldots,n_{m-1}$ be integers with $n_a=1$ for $a\notin D$
and $n_a\in\{0,1,2\}$ for $a\in D$. If
\[
 \sum_a n_a=m,\qquad \sum_a a n_a\equiv\sum_a a\pmod m,
\]
then $n_a=1$ for every $a$.
\end{lemma}
\begin{proof}
Put $\epsilon_a=n_a-1$. Then $\epsilon_a\in\{-1,0,1\}$ on $D$,
$\sum\epsilon_a=0$, and $\sum a\epsilon_a\equiv0\pmod m$.
The last sum has absolute value less than $m$, so it is zero as an
integer. If $a$ were the largest positive label with $\epsilon_a\ne0$, then
$|a\epsilon_a|>\sum_{b<a}|b\epsilon_b|$, contradicting the vanishing
of the weighted sum. The equation $\sum_a\epsilon_a=0$ then gives
$\epsilon_0=0$.
\end{proof}
Choose $D$ satisfying the hypotheses of \cref{finite:lem:marked-counts} and $2^{|D|}\ge |S|$. For example, one can take $0$ and an initial segment of $1,2,4,\ldots$ or $1,3,7,\ldots$ whose length gives enough distinct vectors and whose sum is less than $m$. Assign the shifts $s\in S$ distinct binary vectors $(\alpha_s(a))_{a\in D}$, and put $\alpha_s(a)=0$ for $a\notin D$. For a single shift we may take $D=\varnothing$.

\begin{theorem}[Cyclic lift with marked coordinates]\label{finite:thm:signature-lift}
Let $m\ge3$ be odd, let $L\ge1$ be coprime to $m$, and let $\delta$
be a unit modulo $m$. Suppose the matchings $P_s$ in
\cref{finite:eq:single-sum} are the only perfect matchings of their union.
Choose $D$ and the bits $\alpha_s(a)$ as above. Fix $\beta\in\{0,1\}$ and take the triples
\begin{equation}\label{finite:eq:signature-lift}
 \left((a,x),\ (\lambda a+s,x+\alpha_s(a)+\beta h),\
 (\delta h-(\lambda+1)a-s,x+\alpha_s(a)+(\beta+1)h)\right)
\end{equation}
for $a\in\mathbb Z_m$, $x\in\mathbb Z_L$, $h\in\{0,1\}$, and
$s\in S$. Here $x$ is the sheet index of the first vertex, and the three base coordinates sum to $\delta h$ modulo $m$. For each fixed pair $(h,s)$ the triples form a perfect matching. Their union over all $(h,s)$ is a balanced tripartite $2|S|$-rigid
$3$-graph of part size $mL$ with maximum pair codegree at most two.
\end{theorem}
\begin{proof}
For $L=1$ the result follows from \cref{finite:lem:double-sum}. Suppose $L\ge2$. For fixed $h,s$, the base coordinate in each part of \cref{finite:eq:signature-lift} is a bijective function of $a$. At each $a$, the corresponding sheet index is a translate of $x$, so these triples form a perfect matching. Projection determines the base edge and hence $(h,s)$. The sheet index of the first vertex gives $x$. Thus the lifted hypergraph is simple and the prescribed matchings are edge-disjoint.

Let $P$ be any perfect matching, and let $u$ count its edges with $h=1$.
Summing the three base coordinates over the edges of $P$ gives $m\mid u$. On each edge the third sheet index minus the second equals $h$ modulo $L$. Each sheet index occurs $m$ times in the second part and $m$ times in the third, so the differences sum to zero modulo $L$. Thus $L\mid u$. Coprimality yields $mL\mid u$.
Since $0\le u\le mL$, all edges of $P$ have a common $h$.

Call the $m$ edges of $P$ meeting sheet $y$ of the second part the \emph{slice at $y$}. They cover sheet $y+h$ of the third part exactly once. For each unmarked $a$, the slice contains exactly one first-part vertex with base coordinate $a$, namely $(a,y-\beta h)$. For a marked coordinate $a$, the bit $\alpha_s(a)$ is 0 or 1, so the first-part sheet is either $y-\beta h$ or $y-\beta h-1$. Thus the multiplicity $n_a$ of $a$ in the slice is 1 for unmarked $a$ and at most 2 for marked $a$. Summing the three base coordinates over these $m$ edges gives $m\delta h\equiv0\pmod m$. The second and third base coordinates each run through $\mathbb Z_m$, so $\sum_a a n_a\equiv0\equiv\sum_a a\pmod m$. By \cref{finite:lem:marked-counts}, each base coordinate of the first part occurs exactly once in this slice. The projection of the slice is therefore one of the prescribed base matchings, indexed by $(h,s_y)$ for some $s_y\in S$.

Fix a marked coordinate $a$ and a first-part vertex $(a,z)$. Put $y=z+\beta h$. The vertex is covered in slice $y$ if $\alpha_{s_y}(a)=0$, and in slice $y+1$ if $\alpha_{s_{y+1}}(a)=1$. Exactly one condition holds, so the two bits agree. This holds at every marked coordinate and every $y$. The bit vectors identify the shifts, hence all $s_y$ coincide and $P$ is a prescribed matching.

By \cref{finite:lem:double-sum}, the base hypergraph has maximum pair codegree at most two. The lifts of a single base edge are pairwise vertex-disjoint. Thus for any fixed pair of vertices in distinct parts of the lift, each base edge containing their projections contributes at most one lifted edge containing the pair. Hence the lift also has maximum pair codegree at most two.
\end{proof}

\section{When cyclic lifts are linear}\label{finite:sec:linear-lifts}
\begin{theorem}[Linear cyclic lifts and small-period refinements]
\label{finite:thm:linear-lift}
Under the hypotheses of \cref{finite:thm:signature-lift}, take $\beta=1$.
\begin{enumerate}
\item If $L\ge4$, the lift with all $2|S|$ prescribed matchings is linear.
\item If $L=3$, the full lift is linear whenever
\begin{equation}\label{finite:eq:monotone-code}
 \alpha_{s+\delta}(a)\le\alpha_s(a)
 \quad\text{for every }a\in\mathbb Z_m\text{ and every }s\in S\text{ with }s+\delta\in S.
\end{equation}
\item If $L=2$, delete the matching $(1,s)$ for each $s\in S\cap(S+\delta)$. The remaining hypergraph is linear and has $2|S|-|S\cap(S+\delta)|$ prescribed perfect matchings.
\end{enumerate}
In each case the retained matchings are the only perfect matchings of their union.
\end{theorem}
\begin{proof}
By \cref{finite:thm:signature-lift}, the full lift is rigid, and deleting any of its prescribed perfect matchings preserves rigidity. For a fixed $h$, a pair of base vertices belongs to at most one base edge, whose lifts are pairwise vertex-disjoint. It therefore remains to compare a lifted edge with $h=0$ and one with $h=1$.

For a pair in the first and second parts, its base coordinates determine the same shift $s$ for both values of $h$. The second sheet index minus the first is $\alpha_s(a)$ for $h=0$ and $\alpha_s(a)+1$ for $h=1$. These are distinct modulo $L$ for $L\ge2$. For a pair in the second and third parts, the third sheet index minus the second gives $h$ directly. To test whether edges of the two types can share their first- and third-part vertices, fix both base coordinates and the sheet index of the first vertex. If the shift is $s$ at $h=0$, it is $s+\delta$ at $h=1$. The third sheet index for $h=1$ minus that for $h=0$ is congruent modulo $L$ to the integer
\[
 2+\alpha_{s+\delta}(a)-\alpha_s(a)\in\{1,2,3\}.
\]
None of these three integers is divisible by $L$ when $L\ge4$. For $L=3$, the value 3 would require $\alpha_s(a)=0$ and $\alpha_{s+\delta}(a)=1$, contrary to \cref{finite:eq:monotone-code}. For $L=2$, any shift $s+\delta$ that could give the same third-part sheet index belongs to $S\cap(S+\delta)$, so the matching $(1,s+\delta)$ is deleted.
\end{proof}

For the \ghzdatafile{examples/twelve_linear_on_117.json}{117-path network}, use the 39-path seed with $m=13$, $\lambda=2$, and $\delta=1$. Take $L=3$, $\beta=1$, and $D=\{0,1,2\}$. For $s=0,1,2,4,5,8$, encode $\alpha_s$ by $7,6,4,5,1,3$, respectively, taking $\alpha_s(a)$ as bit $a$. The vectors are distinct. For every $a\in D$, we have $\alpha_0(a)\ge\alpha_1(a)\ge\alpha_2(a)$ and $\alpha_4(a)\ge\alpha_5(a)$. By \cref{finite:thm:signature-lift,finite:thm:linear-lift}, the three-sheet lift has exactly twelve perfect matchings and any two source supports share at most one path. The \ghzdatafile{certificates.md}{parameters of the other lifts} give the binary vectors and the matchings deleted in the two-sheet cases.

\section{Small unrestricted triple systems}\label{finite:sec:finite-inputs}
To check a finite source list, the \ghzdatafile{verification/verify_export.py}{exact-cover program} chooses an uncovered vertex and tries each incident triple disjoint from those already selected. Every completion uses exactly one of these triples, so the search counts each perfect matching once.

\begin{proposition}\label{finite:prop:thirteen18}
There is a simple $13$-rigid $3$-graph on 18 vertices. In particular, $f_3(18)\ge13$.
\end{proposition}
\begin{proof}
Take two copies of the nine paths in \cref{tab:finite-thirteen}. For each $M_i$, the table below lists the set $U_i$ of paths that change layers. The lift of $e\in M_i$ from layer $j$ uses $(v,j)$ for $v\in e\setminus U_i$ and $(v,1-j)$ for $v\in e\cap U_i$. For example, $036\in M_1$ lifts to $\{(0,j),(3,j),(6,1-j)\}$ for $j=0,1$.
\begin{center}
\begin{tabular}{@{}rl@{\qquad}rl@{\qquad}rl@{}}
\toprule
$i$&$U_i$&$i$&$U_i$&$i$&$U_i$\\\midrule
0&$\varnothing$ &5&$\{2,5,6,8\}$ &10&$\{1,3,5\}$\\
1&$\{5,6\}$     &6&$\{3,6,7,8\}$ &11&$\{2,4,6,7,8\}$\\
2&$\{4,6,7\}$   &7&$\{3,5\}$     &12&$\{4,5,6,7\}$\\
3&$\{3,6,7\}$   &8&$\{4,5,8\}$   &&\\
4&$\{8\}$       &9&$\{3,4,7,8\}$ &&\\
\bottomrule
\end{tabular}
\end{center}
The six lifted triples from each $M_i$ cover both copies of every path exactly once. Projecting a lifted triple recovers its base triple, and the two lifts of a single triple are disjoint. Thus the 13 matchings partition 78 distinct triples. With $(v,j)$ numbered $2v+j$, the exact-cover check on the \ghzdatafile{examples/thirteen_on_18.json}{resulting source list} finds exactly these 13 matchings.
\end{proof}

\begin{theorem}\label{finite:thm:ten-all}
For every integer $m\ge2$, we have $f_3(3m)\ge10$. Moreover $f_3(3)=1$, $f_3(6)=10$, and $\min_{m\ge2}f_3(3m)=10$.
\end{theorem}
\begin{proof}
The \ghzdatafile{retained_v11/verification/ten_states_orders_two_to_eleven.json}{ten initial source lists} give examples at every matching size $2\le a\le11$. The \ghzdatafile{retained_v11/verification/verify_small_ten_seeds.py}{finite check} finds exactly ten perfect matchings in each and verifies the tripartition at $a=11$. Grafting this balanced example increases the matching size by 10 and gives another $10$-rigid hypergraph (\cref{finite:thm:graft}). Each $m\ge2$ has the form $a+10q$ for some $2\le a\le11$ and $q\ge0$, so $q$ grafts give the required example. On three vertices there is just one possible triple. On six vertices the complete $3$-graph has exactly ten complementary pairs of triples, proving the two exact values and the claimed minimum.
\end{proof}

\section{Thresholds from the finite seeds}\label{finite:sec:thresholds}

\Cref{finite:tab:allorders} compares the three examples in \cref{tab:twelve-networks} with the thresholds proved below.

\begin{table}[H]
\centering
\caption{Triplet-source networks of dimension 12. For each class we give the path count of one example and a threshold $n_0$. Every $n\ge n_0$ divisible by 3 is also a possible path count in that class.}\label{finite:tab:allorders}
\begin{tabular}{lrrr}
\toprule
Source supports & Paths in the example & Matching size $m_0$ & Paths $n_0=3m_0$\\
\midrule
Unrestricted & 9 & 21 & 63\\
Balanced tripartite & 39 & 43 & 129\\
Linear balanced tripartite & 117 & 113 & 339\\
\bottomrule
\end{tabular}
\end{table}

\begin{theorem}[Balanced tripartite and unrestricted thresholds]\label{finite:thm:thresholds}
For each $t$ in the table below, there is a balanced tripartite $t$-rigid $3$-graph with $m$ vertices per part and maximum pair codegree at most 2 whenever $m\ge m_0$ or $m$ is odd with $m\ge m_1$. Thus $g_3(m)\ge t$ at every such size.
\[
\begin{array}{@{}l*{3}{r}@{}}
 t   &8&10&12\\
 \midrule
 m_0 &7&31&43\\
 m_1 &7&17&35
\end{array}
\]
For unrestricted triples, $f_3(3m)\ge12$ for every $m\ge21$.
\end{theorem}
\begin{proof}
By \cref{finite:eq:graft-f,finite:eq:graft-g}, grafting a balanced tripartite seed of part size $q$ increases the matching size by $q-1$. In each row below we retain $t$ perfect matchings in every seed. Each matching size in the last two columns is obtained by starting at a size in either seed column and adding increments $q-1$ from the grafts.

\begin{center}
\begin{tabular}{@{}lrcccc@{}}
\toprule
Source supports & $t$ & Graft sizes $q$ & Other seed sizes & Odd starts & Even starts\\
\midrule
Tripartite & 8  & $5$ & $7,8,10$ & $7,9$ & $8,10$\\
Tripartite & 10 & $11,13,17,19$ & $22,26$ & $17,19,\ldots,25$ & $32,34,\ldots,40$\\
Tripartite & 12 & $13,19,23,29$ & $26,39,46,52$ & $35,37,\ldots,45$ & $44,46,\ldots,54$\\
Unrestricted & 12 & $13,19,23,29$ & $3,4,6$ & $19,21,\ldots,29$ & $22,24,\ldots,32$\\
\bottomrule
\end{tabular}
\end{center}

The modular construction in \cref{finite:prop:eight-coprime} gives the tripartite seeds of sizes 5 and 7, and the Latin square in \cref{finite:eq:latin-odd} gives size 8. The other tripartite seeds come from \cref{finite:lem:double-sum,finite:thm:signature-lift}. In particular, $10,22,26,46$ are two-sheet lifts of $5,11,13,23$, respectively, and $39,52$ are the three- and four-sheet lifts of the size-13 seed. Each tripartite seed has maximum pair codegree at most 2, a bound preserved by \cref{finite:eq:graft-degree}. For unrestricted triples, the 9-path example in \cref{tab:finite-thirteen} and its 18-path lift in \cref{finite:prop:thirteen18} give 13 matchings at sizes 3 and 6. At size 4 the exact-cover check of the \ghzdatafile{examples/twelve_on_12.json}{12-path source list} finds exactly 12 perfect matchings.

For the tripartite row with $t=12$, the following sums give the starting sizes in increasing order within each parity.
\[
\begin{array}{@{}lrrrrrr@{}}
\text{Odd}  & 13+22 & 19+18 & 39 & 13+28 & 13+12+18 & 23+22\\
\text{Even} & 26+18 & 46 & 26+22 & 26+12+12 & 52 & 26+28
\end{array}
\]
Each list covers all residues of its parity modulo 12. Repeated grafts with the size-13 seed therefore give all odd $m\ge35$ and all even $m\ge44$, hence every $m\ge43$. For the other rows the steps are $4,10,12$, respectively. In each row the listed starts likewise cover every residue of their parity modulo the step. This gives all $m\ge7$ in the first row, all odd $m\ge17$ and even $m\ge32$ in the second, and all odd $m\ge19$ and even $m\ge22$ in the last.
\end{proof}

\subsection{Thresholds for linear tripartite hypergraphs}
\begin{theorem}\label{finite:thm:linear-thresholds}
Let $(t,m_0)$ be one of $(8,33)$, $(9,62)$, $(10,74)$, $(11,112)$, and $(12,113)$. For every $m\ge m_0$ there is a linear balanced tripartite $t$-rigid $3$-graph with $m$ vertices in each part.
\end{theorem}
\begin{proof}
Fix $t$ and retain $t$ matchings in each of the linear examples used for this dimension. These examples come from the Latin square in \cref{finite:sec:periodic} and the affine lifts in \cref{finite:thm:linear-lift}. A seed of part size $q$ contributes the increment $q-1$ under grafting. By \cref{finite:thm:graft}, each nonempty sum of these increments yields a linear example of part size one plus the sum.

The \ghzdatafile{retained_v11/verification/linear_thresholds_refined.json}{residue certificates} specify an available increment $a$ and, for each residue $r$ modulo $a$, a sum $w_r$ of available increments congruent to $r$. The empty sum represents zero. The values of $a$ and $W=\max_r w_r$ are
\[
\begin{array}{@{}lrrrrr@{}}
t & 8 & 9 & 10 & 11 & 12\\
\midrule
a & 7 & 21 & 25 & 37 & 38\\
W & 38 & 81 & 97 & 147 & 149
\end{array}
\]
and the \ghzdatafile{retained_v11/verification/verify_linear_arithmetic.py}{verification program} checks the seeds and each sum. Put $s=m-1$ and choose $r\equiv s\pmod a$. If $s\ge W-a+1$, then $s\ge w_r$, since otherwise $s\le w_r-a\le W-a$. Hence $s=w_r+ja$ for some $j\ge0$. Since $s>0$, this gives a nonempty sum and proves the bound for all $m\ge W-a+2$.

For $t=12$, the 117-path example gives $a=39-1=38$. The largest representative is $w_{35}=149=57+92$, from seeds of part sizes 58 and 93. Thus $m_0=149-38+2=113$. For every $n\ge339$ divisible by 3, there is a linear tripartite network of dimension 12 on $n$ paths.
\end{proof}

\stopcontents[supplement]
\ifstandalonesupplement
  \bibliographystyle{abbrv}
  \bibliography{references}
\fi

\end{document}